%% file: CCSN2022_main.tex
\documentclass[twocolumn, tighten]{aastex631}

\hypersetup{linkcolor=red,citecolor=green,filecolor=cyan,urlcolor=magenta}

\usepackage[latin1]{inputenc}
\usepackage{amsmath}
\usepackage{amsfonts}
\usepackage{amssymb}
\usepackage{units}
\usepackage{nicefrac}
\usepackage{graphicx}
\usepackage{hyperref}
\usepackage{longtable}
\usepackage{xspace}
\usepackage{float}

\newcommand\se{stripped-envelope\xspace}  

\submitjournal{ApJ Letters}

\shorttitle{Neutrinos from Supernovae}
\shortauthors{R. Abbasi et al.}

\begin{document}

\title{Constraining High-Energy Neutrino Emission from Supernovae with IceCube}

\include{authors}



\begin{abstract}

Core-collapse supernovae are a promising potential high-energy neutrino source class. We test for correlation between seven years of IceCube neutrino data and a catalog containing more than 1000 core-collapse supernovae of types IIn and IIP and a sample of \se  supernovae. We search both for neutrino emission from individual supernovae, and for combined emission from the whole supernova sample through a stacking analysis. 

No significant spatial or temporal correlation of neutrinos with the cataloged supernovae was found. The overall deviation of all tested scenarios from the background expectation yields a p-value of $93\%$ which is fully compatible with background. The derived upper limits on the total energy emitted in neutrinos are $1.7\times 10^{48}$\,erg for \se supernovae,  $2.8\times 10^{48}$\,erg for type IIP, and $1.3\times 10^{49}$\,erg for type IIn SNe, the latter disfavouring models with optimistic assumptions for neutrino production in interacting supernovae. 

We conclude that \se supernovae and supernovae of type IIn do not contribute more than $14.6\%$ and $33.9\%$ respectively to the diffuse neutrino flux in the energy range of about $\unit[10^3-10^5]{GeV}$, assuming that the neutrino energy spectrum follows a power-law with an index of $-2.5$. Under the same assumption, we can only constrain the contribution of type IIP SNe to no more than 59.9\%. Thus core-collapse supernovae of types IIn and \se supernovae can both be ruled out as the dominant source of the diffuse neutrino flux under the given assumptions. 

\end{abstract}



\section{Introduction} 
\label{sec:introduction}

IceCube has detected a diffuse flux of high-energy astrophysical neutrinos \citep{Aartsen:2015knd, Aartsen:2013jdh}. 
The majority of the high-energy neutrinos follows an isotropic distribution which suggests an extra-galactic origin.
The active galaxy NGC 1068 was recently reported to be the first extra-galactic point-source of high-energy neutrinos beyond the $4 \sigma$ level \citep{icecubecollaborationEvidenceNeutrinoEmission2022}. 
%
%
While there is evidence that gamma-ray blazars and tidal disruption events (TDEs) produce high-energy neutrinos~\citep{MWScience,ICScience,Stein:2020xhk,reuschCandidateTidalDisruption2022a}, the rate of observed coincidences constrain the overall diffuse flux contribution of resolved gamma-ray blazars and tidal disruption events to no more than $30\%$~\citep{2017ApJ...835...45A} and $26\%$~\citep{Stein:2019lI} respectively, leaving the majority of the diffuse flux unexplained.

In general, high-energy neutrinos are created through interactions of high-energy protons with ambient matter or photon fields. Charged and neutral pions produced in those interactions decay to neutrinos and gamma rays, respectively. While gamma rays can also be produced in leptonic processes such as Inverse Compton scattering, neutrinos are considered to be the clear signature for hadronic interactions, and thus also cosmic-ray acceleration.

Several source classes have been proposed as candidate neutrino (and cosmic-ray) sources. Among the most promising are Active Galactic Nuclei (AGN), Gamma-Ray Bursts (GRBs) and Supernovae (SNe) -- see \citet{Kurahashi:2022utm} for a recent review. While gamma-bright GRBs are strongly disfavored as the main contributor to the measured diffuse neutrino flux \citep{2017ApJ...843..112A}, a large population of nearby low-luminosity bursts could still contribute significantly. The discovery of a connection between GRBs and type Ic-BL SNe implies that (mildly) relativistic jets should also exist in a fraction of core-collapse SNe \citep{Senno:2015tsn, 2005PhRvL..95f1103A,2004PhRvL..93r1101R,2018ApJ...855...37D}, where such jets might be choked inside the envelope of the star. In this scenario, the gamma rays would be absorbed but the neutrinos could still escape. A short neutrino burst ($\sim 100$\,s) would be expected, in coincidence with the explosion time of the SNe \citep{Senno:2015tsn}.

Another possibility for producing high-energy neutrinos in core-collapse supernovae (CCSN) is through interactions of the SN ejecta with a dense circumstellar medium (CSM). 
Strong stellar winds in the star's late evolution stages or pre-outburst could produce a sufficiently-dense CSM \citep{Ofek:2013mea,2021ApJ...907...99S}. When the supernova shock front reaches this dense medium, efficient acceleration of charged particles on timescales ranging from a few tens of seconds to $\sim 1000$ days 
may occur \citep{Murase:2010cu, Zirakashvili:2015mua,sarmahHighEnergyParticles2022}. CSM interactions can be revealed through the detection of a combination of narrow and broad emission lines (as observed in type IIn SNe). The narrow component of the spectral lines is produced by circumstellar gas, which is ionised as the shock breaks out of the star. The intermediate and broad components are produced by shocked, high-velocity SN ejecta, arising as a result of the collision of the ejecta with circumstellar gas. Another indication might be a long plateau in the SN light curve (as seen in Type IIP SNe), which could be partly powered by SN shock breakout interaction with dense CSM~\citep{2011MNRAS.415..199M,2012IAUS..279...54M}. Some IIP SNe show direct observational evidence for interactions~\citep{mauerhan2013, faran2014, Yaron:2017umb, nakaoka2018}. 
\citet{2022ApJ...929..163P} found the high-energy neutrino IC200530A in spatial coincidence with the optical transient AT2019fdr, which they interpret as a Type IIn superluminous supernova.

Optical follow-up campaigns of IceCube high-energy neutrino alerts \citep{steinNeutrinoFollowupZwicky2022, neckerASASSNFollowupIceCube2022} are close to constraining the brightest observed superluminous supernovae. 

Here, for the first time, we probe different SN classes as potential neutrino sources and calculate their possible contribution to the observed diffuse neutrino flux. To search for cross-correlation between neutrinos and optically observed SNe, we utilize data recorded by the IceCube Neutrino Observatory.

This paper is organized as follows: Section \ref{sec:data} describes the relevant data sets, followed by a discussion of the analysis methods in Section \ref{sec:AnalysisMethod} and the presentation of the results in Section \ref{sec:Results}. Section \ref{sec:DiffuseNeutrinoFlux} presents the constraints on the contribution of CCSNe to the diffuse neutrino flux. Section \ref{sec:Conclusion} summarizes the paper. Upper limits on the total energy released in neutrinos from individual SNe can be found in the Appendix \ref{sec:UpperLimits}.

\section{The Data}
\label{sec:data}
IceCube is a cubic-kilometer-sized neutrino detector, located in the transparent ice of the 2.8\,km-thick glacier covering the bedrock at the geographical South Pole~\citep{Aartsen:2016nxy}.
Neutrino-nucleon interactions in the ice are detected indirectly, via Cherenkov light emission from secondary particles, by 5160 photomultiplier tubes. While charged-current interactions of muon neutrinos produce track-like signatures with sub-degree angular resolution, both charged-current interactions of electron and tau neutrinos, and neutral-current interactions, have angular resolutions of several degrees. This analysis utilizes a selection of seven years of IceCube muon-track data that was optimised for point-source searches~\citep{Aartsen:2016oji}, with roughly 700,000 events from years 2008 to 2014.

\begin{figure}
\centering \includegraphics[width=.5\textwidth]{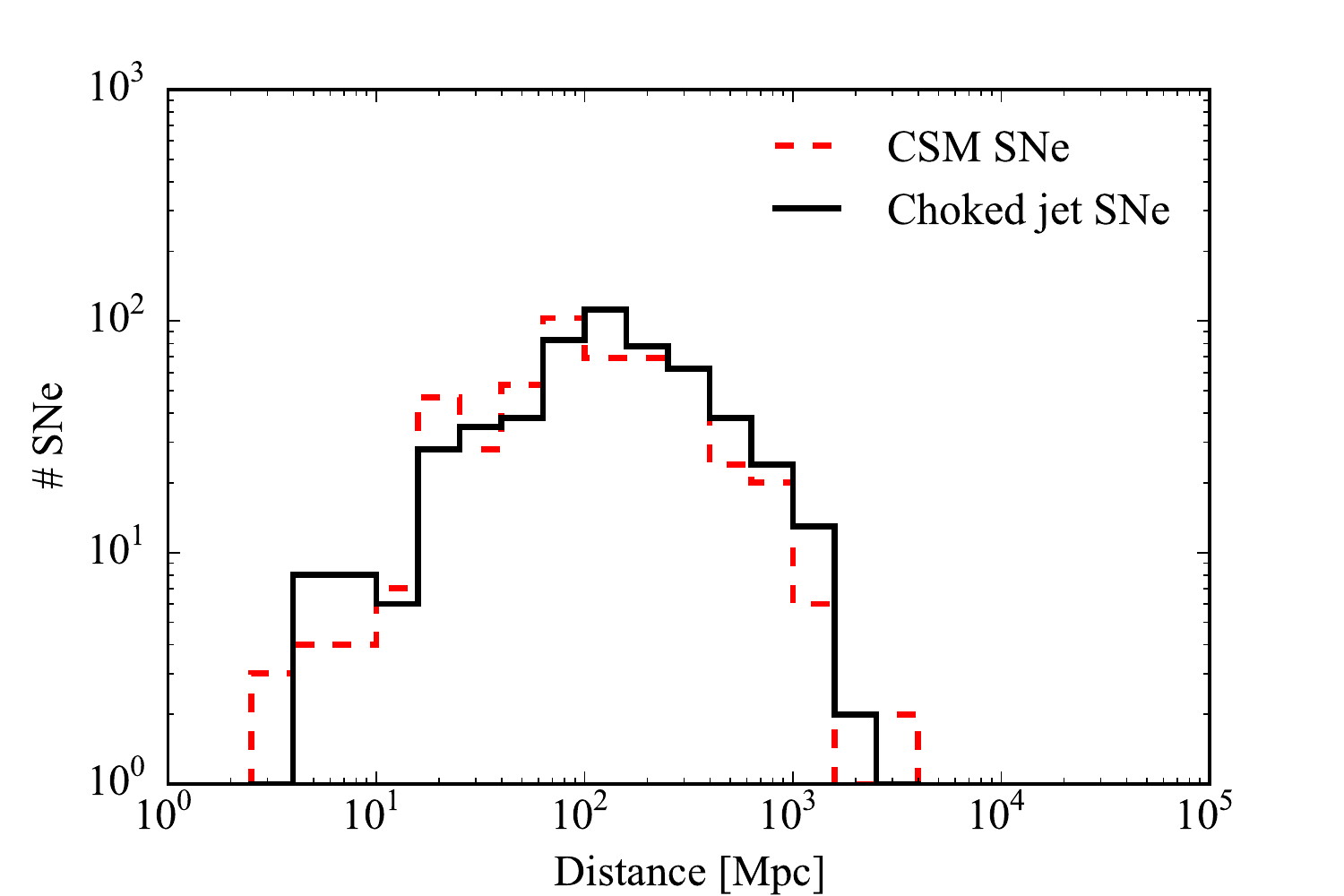}
\caption{Distance distribution of CSM SN sample and the \se SN sample. The decrease at large distance is a result of 
limited detection sensitivity
and a selection bias towards brighter objects, which are easier to classify spectroscopically.}
\label{fig:DistanceDistribution}
\end{figure}

The CCSN catalog for this analysis was compiled using publicly available records of optical detections of SNe. The primary sources were the WiseREP SN catalog~\citep{Yaron:2012aa} and the OpenSupernovaCatalog \citep{Guillochon:2016rhj}. 
In total, the compiled source sample contains 339 SN type IIn, 198 SN type IIP, 503 type Ib/c and type IIb SNe. The latter are referred to as \se supernovae. Both type IIn and type IIP SN are candidates for CSM interaction, while \se supernovae might host choked jets. In Figure~\ref{fig:DistanceDistribution} the distance distribution of the two sub-samples is shown. Note that while we did include many supernovae in the analysis, we list only those of a smaller subsample in Tables \ref{tab:CatalogueIIP}, \ref{tab:CatalogueSE} and \ref{tab:CatalogueIIn} as explained below.

The distance was taken from the previously-cited catalogs. For cases in which entries were missing in the catalogs, the distances were estimated using redshift measurements. The $\Lambda$CDM model, with cosmological parameters measured by Planck \citep{Ade:2015xua}, were used to convert from redshift to luminosity distance. We have assumed a peculiar motion of $\unit[300]{km \, s^{-1}}$, which also provides a lower distance limit for SNe with very small redshifts. SNe with neither distance nor redshift measurements were excluded from the catalog. The distance distribution peaks at about $\unit[100]{Mpc}$, as can be seen in Fig.~\ref{fig:DistanceDistribution}.

\section{Analysis Method}
\label{sec:AnalysisMethod}
To find an excess of neutrinos from the given SN positions and times, a time dependent point source likelihood method \citep{Braun:2009aa} is used. The likelihood function is given by

\begin{equation}
\mathcal{L} = \prod_{i=1}^{N} \left( \frac{n_\text{s}}{N}\mathcal{S}(\nu_i) + \left(1-\frac{n_\text{s}}{N} \right) \mathcal{B}(\nu_i) \right)
\label{eq:Likelihood_general_likelihood}
\end{equation}

where $N$ is the number of neutrino events, $\nu_i$ the $i$th neutrino and $n_\text{s}$ is the number of signal events. $\mathcal{S}$ and $\mathcal{B}$ are signal and background probability distribution functions (PDFs). Each PDF is a product of a spatial term $\mathcal{N}$, an energy term $\mathcal{E}$ and a time term $\mathcal{T}$, which gives for the signal PDF:

\begin{equation}
\mathcal{S} = \mathcal{N}_\mathcal{S} \times \mathcal{E}_\mathcal{S}(\gamma) \times \mathcal{T}_\mathcal{S}
\end{equation}

and similarly for the background PDF $\mathcal{B}$ \citep{Braun:2009aa}. The signal time PDF $\mathcal{T}_\mathcal{S}$ corresponds to the expected neutrino flux as a function of time (light curve) and the background time PDF $\mathcal{T}_\mathcal{B}$ assumes a constant background rate. 

The energy term $\mathcal{E}$ describes the expected neutrino energy spectrum.  As the dataset is highly background-dominated
\footnote{For an astrophysical signal component in the dataset with a spectral index of $\gamma=2.5$, we expect $\mathcal{O}(10^{3})$ signal events and $\mathcal{O}(10^{5})$ atmospheric background events, amounting to a signal contribution of $< 1\%$}
, we can safely assume that the signal contribution is negligible. The background energy proxy distribution is thus assumed to follow the distribution observed in the data. The signal neutrino energy distribution is described as a power-law function, $E^{-\gamma}$, where $\gamma$ is the spectral index. For similar reasons, the background spatial PDF as a function of declination is chosen to match the distribution of declinations found in data. We assume that the background spatial PDF is uniform in right ascension, leading to $\mathcal{N}_\mathcal{B}(\delta, \phi) = \frac{1}{2 \pi} \times \mathcal{N}_{\mathcal{B}, \text{dec}}(\delta)$ for source declination $\delta$ and right ascension $\phi$.  The signal spatial PDF, $\mathcal{N}_\mathcal{S}$, is assumed to follow a 2D-Gaussian distribution.

The likelihood function is maximized with respect to $n_\text{s}$ and $\gamma$. The best fitted value $n_\text{s}$ gives an estimate of the number of signal-like events, i.e., those that are likely to originate from a given SN.

In addition to probing the neutrino fluxes from single SNe, we combine the signal of a sample of SNe with a stacking analysis. Such a source stacking is implemented through a weighted sum of the signal PDFs $\mathcal{S}_{j}$ of individual SNe $j$:

\begin{equation}
\mathcal{S} = \sum_{j}w_j\mathcal{S}_j
\end{equation}

where the weights $w_j$ represent the expected signal strength of the sources. In this analysis, the weights are assumed to be proportional to:

\begin{align}
w_j& \propto \underbrace{\frac{\Phi_0}{D_\text{p}^2}}_{\substack{\text{Source} \\ \text{Properties}} } 
\times \underbrace{\int_{t_\text{start}}^{t_\text{end}} \int_{E_\text{min}}^{E_\text{max}} \, L_j^\nu \, E^{-\gamma}\, A_{\mathrm{eff}}\,\mathrm{d}t \mathrm{d}E}_{\text{Time Dependence}}
\label{eq:GeneralWeightingPrescriptionAlsoNExp}
\end{align}

with $\Phi_0$ as the intrinsic neutrino power of the sources, $D_\text{p}$ as the proper distance \citep{Hogg:1999ad} of the SN, $L_j^\nu(t)$ the estimated neutrino light curve, $E^{-\gamma}$ the neutrino energy spectrum and $A_{\mathrm{eff}}(t, \delta_j, E)$ the effective area, the energy $E$ and the declination of the source $\delta$. The effective area is time-dependent, because the dataset covers several distinct phases of detector construction. The weighting scheme assumes a standard candle ansatz, since we assume the same $\Phi_0$ for each source. It is very sensitive to the estimated source distances, which can have large uncertainties.

A more detailed investigation of the Supernova light curves could mitigate these uncertainties but the optical lightcurves of the supernovae in our catalogue are typically sparse and make detailed modeling complicated. Wrongly estimated weights will impact the sensitivity of the analysis so for the first time in an IceCube analysis we use a novel method of directly fitting the weights $w_{j}$. Adding the flux per source as an additional free parameter 
to the maximum likelihood removes the standard candle assumption and also the dependence on the SN distance estimate, but requires a more advanced numerical procedure to maximize the likelihood function. To test the power of this method, we simulated five sources with random positions on the sky and respective weights. We then perturbed the weights according to a log-normal distribution and used them to compute the sensitivity of the standard, fixed-weights likelihood. Comparing to this, we find an improvement of up to 40\% when using the fitting-weights likelihood. 
We applied this method in addition to the traditional standard-candle one, yielding two separate results. 
%
%

We define the test statistic (TS) by
\begin{equation}
\lambda = 2\times \log \left( \frac{\mathcal{L}(\hat{n}_\text{s}, \hat{\gamma})}{\mathcal{L}(0)} \right)
\end{equation}
where $\mathcal{L}(\hat{n}_\text{s}, \hat{\gamma})$ corresponds to the maximum of the likelihood function and $\mathcal{L}(0)$ to the null hypothesis, i.e., the case of neither spatial nor temporal correlation of neutrinos and SNe~\citep{Braun:2008bg,Braun:2009aa}. The TS distribution is estimated by generating background-only datasets and maximizing the likelihood function with respect to $n_\text{s}$ and $\gamma$. Repeating this procedure many times gives a numerical estimate of the TS distribution. Given an experimental outcome $\lambda_\text{exp}$ and the TS distribution $P(\lambda)$, the p-value is computed as $p = \int_{\lambda_\text{exp}}^\infty P(\lambda)\mathrm{d}\lambda$.

\section{Constraints on supernova subclasses}
\label{sec:Results}
%
%

In the following we present results for selected individual CCSNe, as well as for different subclasses of CCSNe. 

Stripped-envelope SNe, which might have choked jets, are expected to emit a short burst of neutrinos in coincidence with the SN explosion time \citep{Senno:2015tsn}. Motivated by theoretical uncertainties in the duration of the expected neutrino emission, and even larger uncertainties in the SNe explosion time due to sparse optical light curve data, we used a box function starting at 20 days before and extending up to the first available optical data. This ensures the inclusion of the explosion time for a typical SN even if the first detection happened at peak time.

All SN types were tested with box function PDFs of length $100$, $300$ and $1000$ days, starting at the first available optical data, since longer neutrino emission would be expected under the scenario of CSM interaction. In addition, for SNe IIn and IIP light curves were tested of the form:

\begin{equation}
\mathcal{T}(t) \propto \left(1+\nicefrac{t}{t_\text{pp}} \right)^{-1}
\end{equation}
where values of $0.02$, $0.2$ and $2$ years were used for the characteristic time scale constant $t_\text{pp}$, as proposed by ~\citet{Zirakashvili:2015mua}. 

We first applied the maximum likelihood method described above to a selection of individual SNe, which were identified based on their expected relative signal strength $w_j$ as promising. We did not find a statistically significant excess for any of the selected sources. 

The resulting upper limits on the total energy emitted in neutrinos 
between $\unit[10^2]{GeV}$ and $\unit[10^7]{GeV}$, assuming an $E^{-2}$ power-law spectrum, are presented in Appendix \ref{sec:UpperLimits}. In the conversion from the number of neutrino events to flux, the systemic uncertainty is estimated to be about $11\%$, mainly arising from uncertainties in the optical properties of the ice and detector effects~\citep{Coenders2016}.

The individual upper limits range from $10^{49}$ to $6.5\times 10^{50}$\,erg\footnote{Calculated by integrating over time.}, which corresponds to 1-65\% of the typical bolometric electromagnetic energy released in SNe. As the individual \se and IIP SNe are typically closer than the IIn, we generally obtain more stringent limits for these objects.
%
%

In order to improve our sensitivity we performed a stacking analysis, looking for a combined excess from a catalogue instead of individual sources. As explained above we separate supernovae into types IIn, IIP and \se SNe. Note that we decided to treat the types IIn and IIP separately because the presence of CSM interaction in IIP is less certain.

Each of the three sub-catalogs was split into two samples, a bright sample of nearby sources, containing about $70\%$
of the expected signal, and a larger sample, containing the remaining dimmer sources. The bright samples include about 10 SN each, depending on the SN class and the model. The catalogues of the bright samples are listed in Tables \ref{tab:CatalogueIIP}, \ref{tab:CatalogueSE} and \ref{tab:CatalogueIIn}. Testing both independent samples allowed us to benefit from the better optical observations of the nearby sources in the small sample, but also utilize the larger statistics in the large sample. 
Because each source adds a free parameter in the likelihood maximisation when fitting the weights, this was only feasible for the smaller bright sample. This sample contains $\mathcal{O}(10)$ sources which  is a manageable amount of fit parameters. For the large sample the standard candle \textit{ansatz} was applied instead.

The p-values are given in Appendix \ref{sec:pvalues}. The most significant pre-trial p-value is $0.62\%$, and is found in the search for neutrinos from the large sample of type IIP SNe in a 1000-day-long box-shaped light curve. This however corresponds to a post-trial p-value  of $19.5\%$, after accounting for the multiple tested scenarios through simulated pseudo-experiments of the ensemble of p-values, and is thus consistent with background expectations. If this excess were due to astrophysical neutrinos, one would expect a corresponding excess in the sample of nearby type IIP SNe, where we do not find such an excess. The second smallest p-value of $6.3\%$ is found for the nearby type IIn SNe in case of the fitted weights for the box-shaped lightcurve model. The overall deviation of all tested scenarios from the background expectation using a Kolmogorov-Smirnov-test leads to a p-value of $29\%$.

\begin{figure}
\centering 
\includegraphics[width=.48\textwidth]{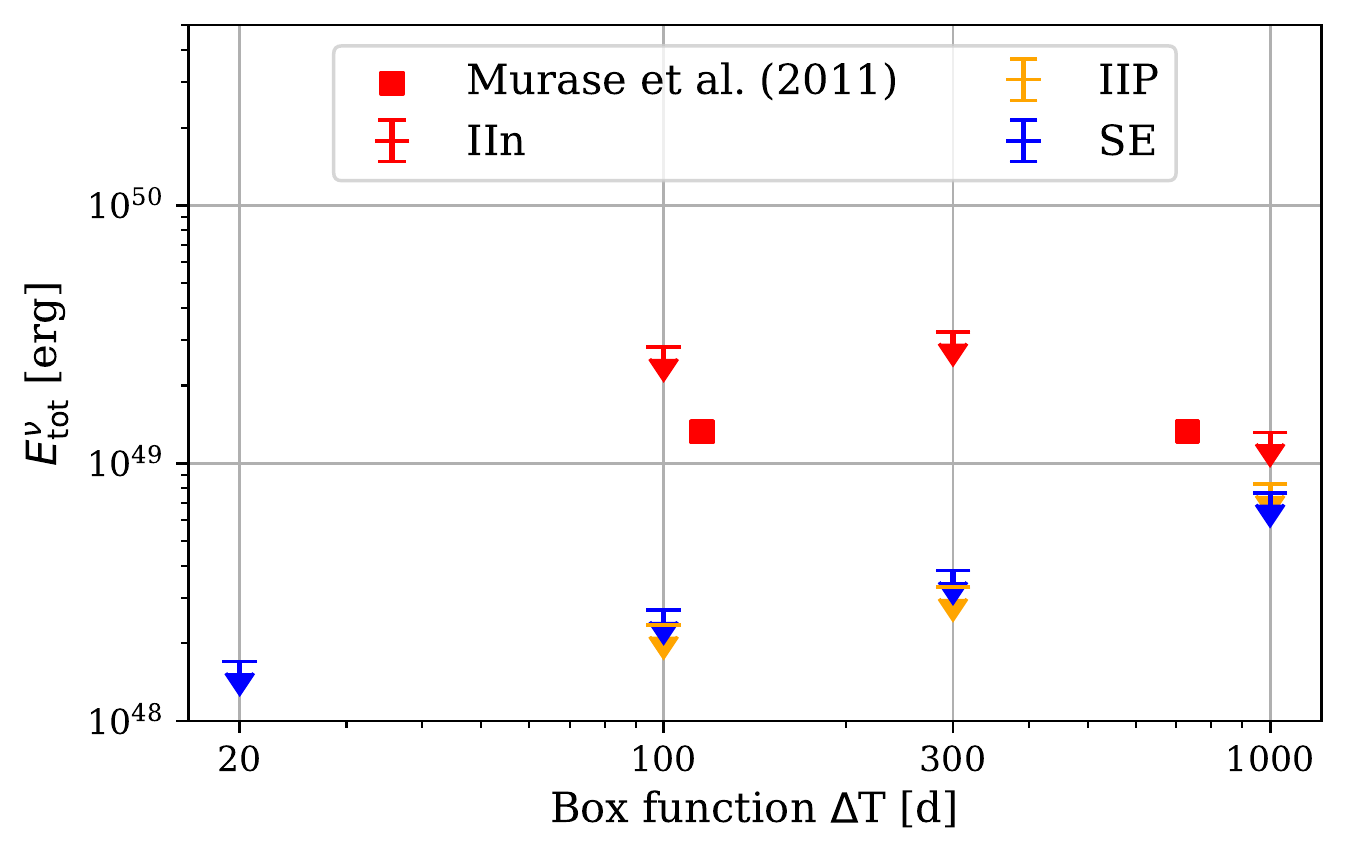}
\includegraphics[width=.48\textwidth]{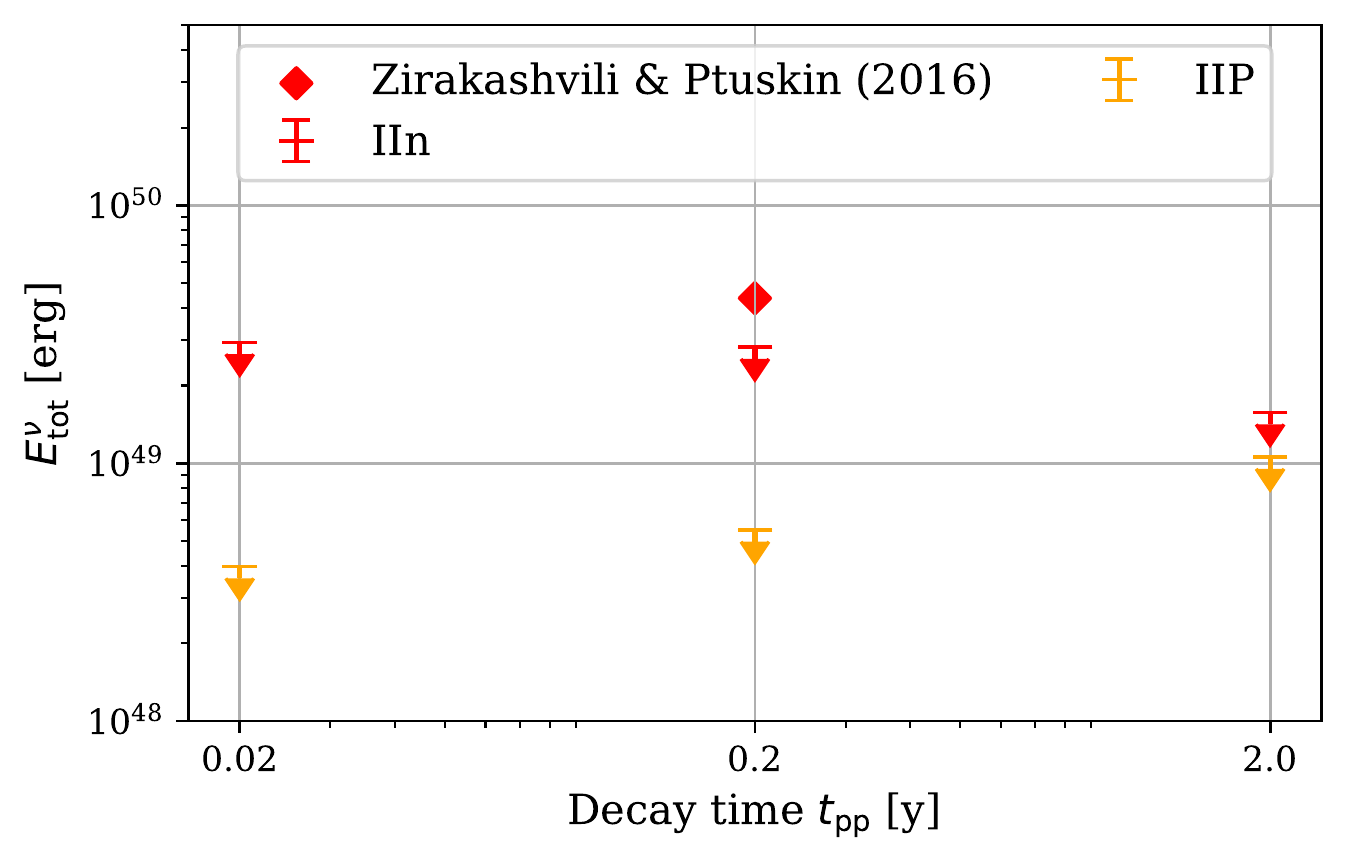}
\caption{Upper limits on total neutrino ($\nu_\mu + \bar{\nu}_\mu$) energy assuming a box-like neutrino light curve (upper panel) and assuming a $L_{\nu}\propto(1+\nicefrac{t}{t_\text{pp}})^{-1}$ neutrino light curve as predicted by~\citet{Zirakashvili:2015mua}. The energy ranges are the same as indicated in Figure \ref{fig:DiffuseFlux}. The model predictions by \citet{Murase:2010cu} and \citet{Zirakashvili:2015mua} are shown as red squares for comparison.}
\label{fig:SensModelsBoxSmall}
\end{figure}

To be conservative we use the result from the fitting weights analysis in the rest of the paper as it resulted in weaker upper limits on the total emitted neutrino energy.
Including systematic uncertainties, those are shown in Fig.~\ref{fig:SensModelsBoxSmall}  for both models of the neutrino light curve. These limits assume that SNe within each category behave as neutrino standard candles. 

The stacking result provides us with stronger limits than individual source limits. We find that SNe type IIn emit less than $1.3\times 10^{49}$\,erg and type IIP less than $2.4\times 10^{48}$\,erg, while the strongest limits for \se SNe of $4.5\times 10^{48}$\,erg are obtained from the ckoked-jet scenario. If the longer box models that are associated with CSM interaction are assumed, then the strongest limit becomes $2.7\times 10^{48}$\,erg.
In general, the box time window provides tighter constraints for CSM interacting SNe compared to the specific light curve model by~\citet{Zirakashvili:2015mua}.

\section{Diffuse Neutrino Flux}
\label{sec:DiffuseNeutrinoFlux}

Using the limits on neutrino energy obtained in the stacking analysis (shown in Fig.~\ref{fig:SensModelsBoxSmall}) we can estimate the maximal contribution from the entire cosmological population of SNe to the measured diffuse neutrino flux~\citep{Aartsen:2015knd}. Using the CCSNe rate by \citet{Strolger:2015kra}, $\rho(z)$, the diffuse flux is computed following the procedure in \citet{Ahlers:2014ioa} assuming a 1:1:1 ($\nu_{e}:\nu_{\mu}:\nu_{\tau}$) neutrino flavour ratio at Earth. Note that we assume that the rate for the individual subclasses scales according to the corresponding percentage in the local universe \citep{liNearbySupernovaRates2011}.
The diffuse flux is given by
\begin{equation}
\phi(E) = \int_0^{\infty} \frac{\rho(z)}{1+z} \frac{dN}{dE} \frac{c}{H(z)} dz,
\end{equation}
where $dN/dE$ is the time integrated flux upper limit for each SN subclass, assuming that the subclass behaves as a neutrino standard candle class with an $E^{-2.5}$ energy spectrum as motivated by the central value of the global fit diffuse neutrino flux \citep{Aartsen:2015knd} and that the power law holds over our sensitive energy range.
This energy range is calculated by finding the energy bound for selecting simulated signal events. We find the values where our sensitivity drops by 5\% for the lower and upper bounds separately. The range between both values is our 90\% energy range.

\begin{figure}
\centering \includegraphics[width=.48\textwidth]{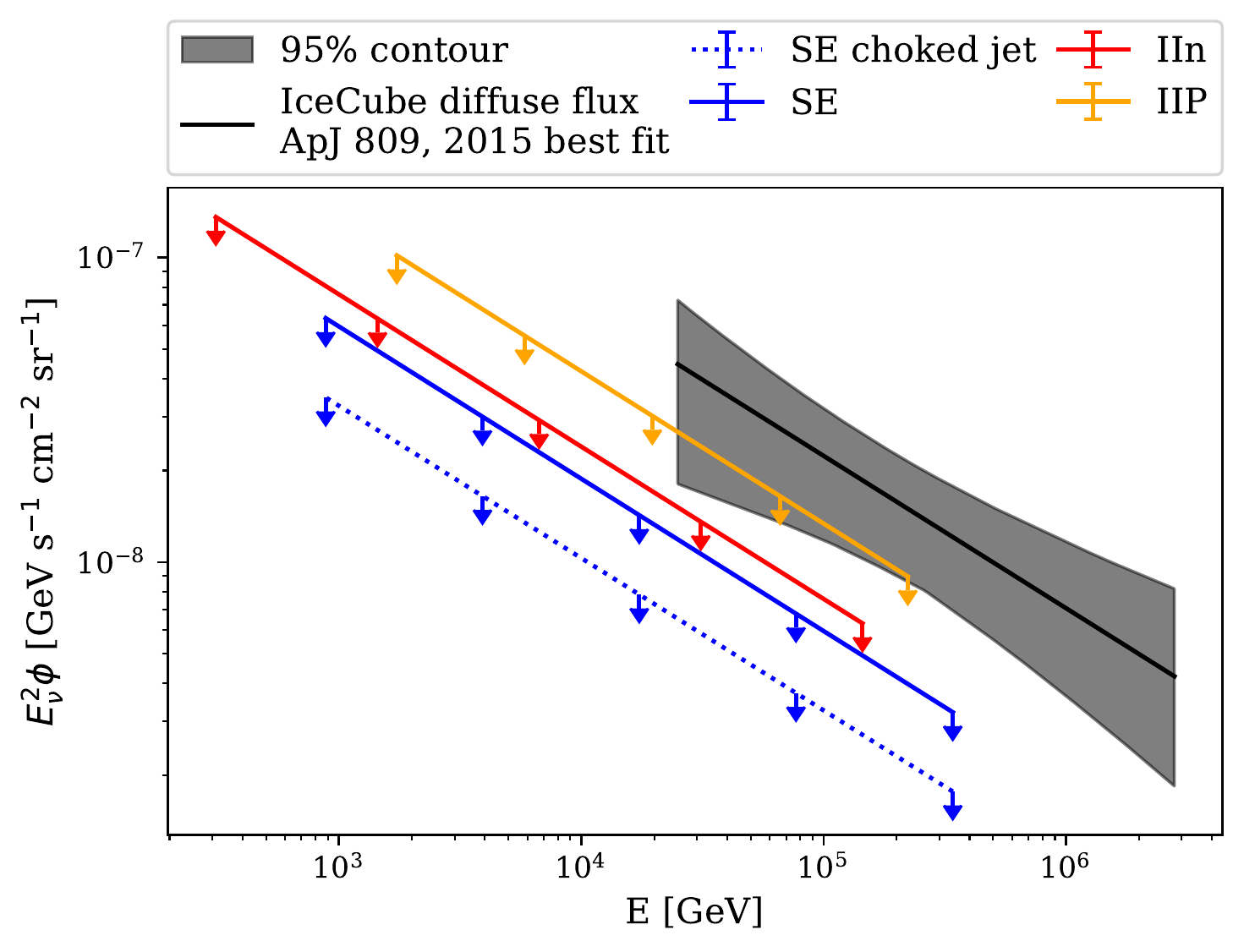}
\caption{
Upper limit on the contribution of different SN types to the diffuse neutrino flux ($\nu_\mu + \bar{\nu}_\mu$) assuming an $E^{-2.5}$ energy spectrum compared with the measured diffuse astrophysical neutrino flux (gray band). The limits are derived from the corresponding strictest limit in Figure \ref{fig:SensModelsBoxSmall}. The choked-jet model refers to the 20-day box model as explained in Section \ref{sec:Results}. The energy range plotted here is the central $90\%$ energy range of the analyzed neutrino sample.} 
\label{fig:DiffuseFlux}
\end{figure}

The resulting upper limits on the contribution of different SN types to the diffuse neutrino flux are shown in Fig.~\ref{fig:DiffuseFlux}.  
Assuming the choked-jet scenario, \se SNe can not contribute more than $14.6\%$ of the observed diffuse neutrino flux. Assuming interaction with the CSM, \se SNe and SNe type IIn can explain not more than $26.6\%$ and $33.9\%$, respectively. We mildly constrain the contribution of SNe type IIP to be less than $59.9\%$. Note that the limit for type IIP SNe seem weaker when translating it to a component of the diffuse flux because they are the most abundant supernova type \citep{liNearbySupernovaRates2011}.
%
%

\begin{figure}[!ht]
\centering \includegraphics[width=.48\textwidth]{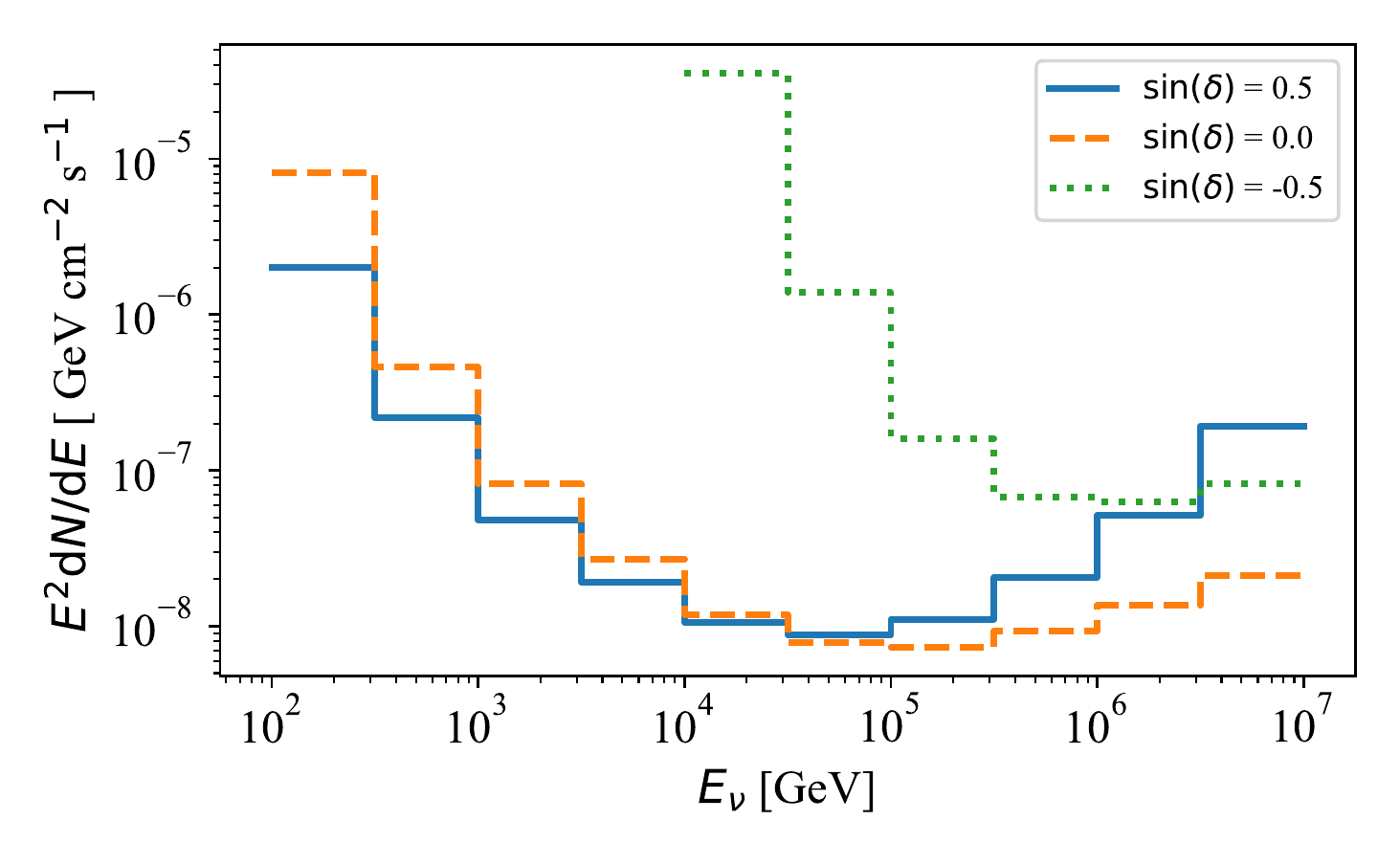}
\caption{Differential sensitivity as a function of energy for different source declinations $\delta$ with one year of experimental data. One can see the maximum sensitivity is achieved around $\unit[10^5]{GeV}$ for sources located in the northern sky and close to the equator. For sources located in the southern sky, the overall sensitivity is much worse, but also peaks at higher energies of $\unit[10^6]{GeV}$.} 
\label{fig:EnergyDist}
\end{figure}

This analysis has different sensitivities for different energy ranges, see Fig. \ref{fig:EnergyDist}. The region of greatest sensitivity is around 10-100 TeV. It can reach to higher energies as well, depending on the source declination. This broadly overlaps with the energy range in which the diffuse IceCube neutrino flux global fit was measured. The quoted upper limits to the diffuse flux contribution are thus not strongly dependent on the extrapolation of the measured diffuse flux to lower energies, where the flux has not yet been measured due to large atmospheric background.

\section{Conclusion}
\label{sec:Conclusion}
We have presented a search for neutrinos from certain types of CCSNe with IceCube. In a stacking analysis we correlated more than 1000 SNe from optical surveys with roughly 700,000 muon-track events recorded by IceCube. The standard stacking method was extended to allow for fitting of individual weights for each source, in order to account for expected variation in the neutrino flux from individual sources.
SNe type IIn, IIP and \se SNe were tested individually with various neutrino emission time models. No significant temporal and spatial correlation of neutrinos and the cataloged SNe was found, allowing us to set upper limits on the contribution of those SNe to the diffuse neutrino flux. 

CCSNe of type IIn, IIP and \se SNe contribute less than $34\%$, $60\%$ and $27\%$ ,respectively, to the diffuse neutrino flux at the $90\%$ confidence level, assuming CSM interaction and an extrapolation of the diffuse neutrino spectrum to low energies following an unbroken power law with index -2.5. Assuming a choked-jet, \se SNe can not contribute more than $15\%$.

Upper limits on the total neutrino energy emitted by a single CSM interacting source are at levels comparable to model predictions by~\citet{Murase:2010cu} (see Fig. ~\ref{fig:SensModelsBoxSmall}) while model predictions from \citet{Zirakashvili:2015mua} are strongly disfavored. Note that the model prediction could easily be adjusted to lower neutrino flux predictions by assuming a lower CSM density or a lower kinetic SN energy. 


Improvements to the presented limits are expected in the near future with optical survey instruments such as the Zwicky Transient Factory \citep{2019arXiv190201945G} which is able to undertake a high-cadence survey across a large fraction of the sky, providing SN catalogs with much greater completeness. In combination with next-generation neutrino telescopes, this will significantly boost the sensitivity of this type of analysis, allowing us to probe dimmer neutrino emitters and smaller contributions of CCSNe to the diffuse neutrino flux.



\section*{Acknowledgments}
The IceCube Collaboration designed, constructed
and now operates the IceCube Neutrino Observatory. Data processing and calibration, Monte Carlo simulations of the detector and of theoretical
models, and data analyses were performed by a large number of collaboration members, who also discussed and approved the scientific results presented here. 
The IceCube collaboration acknowledges the significant contributions to this manuscript from Jannis Necker, Alexander Stasik and Robert Stein. 
It was reviewed by the entire collaboration before publication, and all authors approved the final version of the manuscript.
We acknowledge support from the following agencies:

USA {\textendash} U.S. National Science Foundation-Office of Polar Programs,
U.S. National Science Foundation-Physics Division,
U.S. National Science Foundation-EPSCoR,
Wisconsin Alumni Research Foundation,
Center for High Throughput Computing (CHTC) at the University of Wisconsin{\textendash}Madison,
Open Science Grid (OSG),
Advanced Cyberinfrastructure Coordination Ecosystem: Services {\&} Support (ACCESS),
Frontera computing project at the Texas Advanced Computing Center,
U.S. Department of Energy-National Energy Research Scientific Computing Center,
Particle astrophysics research computing center at the University of Maryland,
Institute for Cyber-Enabled Research at Michigan State University,
and Astroparticle physics computational facility at Marquette University;
Belgium {\textendash} Funds for Scientific Research (FRS-FNRS and FWO),
FWO Odysseus and Big Science programmes,
and Belgian Federal Science Policy Office (Belspo);
Germany {\textendash} Bundesministerium f{\"u}r Bildung und Forschung (BMBF),
Deutsche Forschungsgemeinschaft (DFG),
Helmholtz Alliance for Astroparticle Physics (HAP),
Initiative and Networking Fund of the Helmholtz Association,
Deutsches Elektronen Synchrotron (DESY),
and High Performance Computing cluster of the RWTH Aachen;
Sweden {\textendash} Swedish Research Council,
Swedish Polar Research Secretariat,
Swedish National Infrastructure for Computing (SNIC),
and Knut and Alice Wallenberg Foundation;
European Union {\textendash} EGI Advanced Computing for research;
Australia {\textendash} Australian Research Council;
Canada {\textendash} Natural Sciences and Engineering Research Council of Canada,
Calcul Qu{\'e}bec, Compute Ontario, Canada Foundation for Innovation, WestGrid, and Compute Canada;
Denmark {\textendash} Villum Fonden, Carlsberg Foundation, and European Commission;
New Zealand {\textendash} Marsden Fund;
Japan {\textendash} Japan Society for Promotion of Science (JSPS)
and Institute for Global Prominent Research (IGPR) of Chiba University;
Korea {\textendash} National Research Foundation of Korea (NRF);
Switzerland {\textendash} Swiss National Science Foundation (SNSF);
United Kingdom {\textendash} Department of Physics, University of Oxford.

%

\vspace{5mm}
\facilities{HST(STIS), Swift(XRT and UVOT), AAVSO, CTIO:1.3m,
CTIO:1.5m,CXO}


\software{flarestack \citep{flarestack}}



\appendix


\section{Catalogs}





\begin{table}[H]
\begin{center}
\begin{tabular}{c|r r c r r c}
\textbf{Name} & \textbf{RA} & \textbf{Dec} & \textbf{Discovery Date} & \textbf{Redshift} & \textbf{Distance} & \textbf{Source}\\ 
& [rad] & [rad] & & & [Mpc] & \\\hline
\input{IIn_catalogue}
\end{tabular}
\end{center}
\caption{Interacting supernovae catalogue. References: \input{IIn_reference_map}}
\label{tab:CatalogueIIn}
\end{table}

\begin{table}[H]
\begin{center}
\begin{tabular}{c|r r c r r c}
\textbf{Name} & \textbf{RA} & \textbf{Dec} & \textbf{Discovery Date} & \textbf{Redshift} & \textbf{Distance} & \textbf{Source}\\ 
& [rad] & [rad] & & & [Mpc] & \\\hline
\input{IIP_catalogue}
\end{tabular}
\end{center}
\caption{IIP catalog. References: \input{IIP_reference_map}} 
\label{tab:CatalogueIIP}
\end{table}

\begin{table}[H]
\begin{center}
\begin{tabular}{c|r r c r r c}
\textbf{Name} & \textbf{RA} & \textbf{Dec} & \textbf{Discovery Date} & \textbf{Redshift} & \textbf{Distance} & \textbf{Source}\\ 
& [rad] & [rad] & & & [Mpc] & \\\hline
\input{Ibc_catalogue}
\end{tabular}
\end{center}
\caption{Stripped-envelope supernovae catalogue. References: \input{Ibc_reference_map}}
\label{tab:CatalogueSE}
\end{table}

\section{p-values [\%]}
\label{sec:pvalues}
\begin{table}[H]
\begin{center}
\begin{tabular}{l|rrrr|rrr}
& \multicolumn{4}{c|}{Box length [days]} & \multicolumn{3}{c}{$t_{pp}$ [years]} \\
& [-20, 0] & [0, 100] & [0, 300] & [0, 10000] & 0.02 & 0.2 & 2.0\\
\hline
\input{p-values}
\end{tabular}
\end{center}
\caption{Pre-trial p-values of the fitted weights scenario given as percentages for a box shaped light curve model of different length and for the CSM model of~\citet{Zirakashvili:2015mua} for different choices of $t_{pp}$.
} 
\label{tab:UnblindingResultsPreTrial}
\end{table}

\section{Upper Limits on Individual Sources}
\label{sec:UpperLimits}
This section shows upper limits on individual SNe. Sources were selected based on their expected neutrino signal. Here we assume a generic neutrino energy spectrum of $E^{-2}$ rather than tying them to the observed diffuse spectral shape and an emission time window of 100 days.
\begin{table}[H]
\begin{center}
\begin{tabular}{c|c|c|c|c|c}
Name & ra & dec & discovery date & distance & Energy Upper Limit\\ 
& [rad] & [rad] & & [Mpc] & $[\unit[10^{49}]{erg}]$\\\hline
CSS140111:060437-123740 & 1.59 & -0.22 & 2013-12-24 & 31.8 & 49.8 \\
PSN J13522411+3941286 & 3.63 & 0.693 & 2015-01-09 & 32.1 & 16.8 \\
PSN J14041297-0938168 & 3.68 & -0.168 & 2013-12-20 & 12.5 & 4.8 \\
PTF10aaxf & 2.54 & 0.166 & 2010-11-03 & 52.3 & 29.5 \\
SN2008S & 5.39 & 1.049 & 2008-02-01 & 5.6 & 5.3 \\
SN2009kr & 1.36 & -0.274 & 2009-11-06 & 16.0 & 19.1 \\
SN2011an & 2.09 & 0.287 & 2011-03-01 & 73.0 & 65.3 \\
SN2011ht & 2.65 & 0.905 & 2011-09-29 & 19.2 & 6.6 \\
SN2012ab & 3.24 & 0.098 & 2012-01-31 & 81.0 & 64.18 \\
SN2013gc & 2.13 & -0.49 & 2013-11-07 & 15.1 & 28.4
\end{tabular}
\end{center}
\caption{Upper limits on selected SNe type IIn.} 
\label{}
\end{table}

\begin{table}[H]
\begin{center}
\begin{tabular}{c|c|c|c|c|c}
Name & ra & dec & discovery date & distance & Energy Upper Limit\\ 
& [rad] & [rad] & & [Mpc] & $[\unit[10^{49}]{erg}]$\\\hline
iPTF13aaz & 2.96 & 0.228 & 2013-03-21 & 16.4 & 1.0\\
SN2012A & 2.73 & 0.299 & 2012-01-07 & 9.0 & 1.0\\
SN2012aw & 2.81 & 0.204 & 2012-03-16 & 9.6 & 1.0\\
SN2014bc & 3.22 & 0.826 & 2014-05-19 & 7.6 & 3.0
\end{tabular}
\end{center}
\caption{Upper limits on selected SNe type IIP.} 
\end{table}

\begin{table}[H]
\begin{center}
\begin{tabular}{c|c|c|c|c|c}
Name & ra & dec & discovery date & distance & Energy Upper Limit\\ 
& [rad] & [rad] & & [Mpc] & $[\unit[10^{49}]{erg}]$\\\hline
iPTF13bvn & 3.93 & 0.033 & 2013-06-17 & 25.8 & 4.0 \\
MASTER OT J120451.50 & 3.16 & 0.471 & 2014-10-28 & 15.0 & 1.0 \\
PTF11eon & 3.53 & 0.823 & 2011-06-01 & 8.0 & 1.1\\
SN2008ax & 3.28 & 0.727 & 2008-03-03 & 5.1 & 1.6 \\
SN2008dv & 0.95 & 1.267 & 2008-07-01 & 10.6& 1.2 \\
SN2010br & 3.16 & 0.777 & 2010-04-10 & 9.9 & 4.1 \\
SN2011jm & 3.38 & 0.046 & 2011-12-24 & 14.0 & 1.8 \\
SN2012cw & 2.68 & 0.06 & 2012-06-14 & 31.3 & 4.3 \\
SN2012fh & 2.81 & 0.434 & 2012-10-18 & 8.6 & 1.1 \\
SN2013df & 3.26 & 0.545 & 2013-06-07 & 10.6 & 1.7 \\
SN2014C & 5.92 & 0.601 & 2014-01-05 & 12.1 & 2.3
\end{tabular}
\end{center}
\caption{Upper limits on selected \se SNe (Ib/c and IIb).} 
\end{table}
\clearpage


\bibliography{CCSN2022_lit, sn_references}{}
\bibliographystyle{aasjournal}



\end{document}

%% file: authors.tex
\affiliation{III. Physikalisches Institut, RWTH Aachen University, D-52056 Aachen, Germany}
\affiliation{Department of Physics, University of Adelaide, Adelaide, 5005, Australia}
\affiliation{Dept. of Physics and Astronomy, University of Alaska Anchorage, 3211 Providence Dr., Anchorage, AK 99508, USA}
\affiliation{Dept. of Physics, University of Texas at Arlington, 502 Yates St., Science Hall Rm 108, Box 19059, Arlington, TX 76019, USA}
\affiliation{CTSPS, Clark-Atlanta University, Atlanta, GA 30314, USA}
\affiliation{School of Physics and Center for Relativistic Astrophysics, Georgia Institute of Technology, Atlanta, GA 30332, USA}
\affiliation{Dept. of Physics, Southern University, Baton Rouge, LA 70813, USA}
\affiliation{Dept. of Physics, University of California, Berkeley, CA 94720, USA}
\affiliation{Lawrence Berkeley National Laboratory, Berkeley, CA 94720, USA}
\affiliation{Institut f{\"u}r Physik, Humboldt-Universit{\"a}t zu Berlin, D-12489 Berlin, Germany}
\affiliation{Fakult{\"a}t f{\"u}r Physik {\&} Astronomie, Ruhr-Universit{\"a}t Bochum, D-44780 Bochum, Germany}
\affiliation{Universit{\'e} Libre de Bruxelles, Science Faculty CP230, B-1050 Brussels, Belgium}
\affiliation{Vrije Universiteit Brussel (VUB), Dienst ELEM, B-1050 Brussels, Belgium}
\affiliation{Department of Physics and Laboratory for Particle Physics and Cosmology, Harvard University, Cambridge, MA 02138, USA}
\affiliation{Dept. of Physics, Massachusetts Institute of Technology, Cambridge, MA 02139, USA}
\affiliation{Dept. of Physics and The International Center for Hadron Astrophysics, Chiba University, Chiba 263-8522, Japan}
\affiliation{Department of Physics, Loyola University Chicago, Chicago, IL 60660, USA}
\affiliation{Dept. of Physics and Astronomy, University of Canterbury, Private Bag 4800, Christchurch, New Zealand}
\affiliation{Dept. of Physics, University of Maryland, College Park, MD 20742, USA}
\affiliation{Dept. of Astronomy, Ohio State University, Columbus, OH 43210, USA}
\affiliation{Dept. of Physics and Center for Cosmology and Astro-Particle Physics, Ohio State University, Columbus, OH 43210, USA}
\affiliation{Niels Bohr Institute, University of Copenhagen, DK-2100 Copenhagen, Denmark}
\affiliation{Dept. of Physics, TU Dortmund University, D-44221 Dortmund, Germany}
\affiliation{Dept. of Physics and Astronomy, Michigan State University, East Lansing, MI 48824, USA}
\affiliation{Dept. of Physics, University of Alberta, Edmonton, Alberta, Canada T6G 2E1}
\affiliation{Erlangen Centre for Astroparticle Physics, Friedrich-Alexander-Universit{\"a}t Erlangen-N{\"u}rnberg, D-91058 Erlangen, Germany}
\affiliation{Physik-department, Technische Universit{\"a}t M{\"u}nchen, D-85748 Garching, Germany}
\affiliation{D{\'e}partement de physique nucl{\'e}aire et corpusculaire, Universit{\'e} de Gen{\`e}ve, CH-1211 Gen{\`e}ve, Switzerland}
\affiliation{Dept. of Physics and Astronomy, University of Gent, B-9000 Gent, Belgium}
\affiliation{Dept. of Physics and Astronomy, University of California, Irvine, CA 92697, USA}
\affiliation{Karlsruhe Institute of Technology, Institute for Astroparticle Physics, D-76021 Karlsruhe, Germany }
\affiliation{Karlsruhe Institute of Technology, Institute of Experimental Particle Physics, D-76021 Karlsruhe, Germany }
\affiliation{Dept. of Physics, Engineering Physics, and Astronomy, Queen's University, Kingston, ON K7L 3N6, Canada}
\affiliation{Department of Physics {\&} Astronomy, University of Nevada, Las Vegas, NV, 89154, USA}
\affiliation{Nevada Center for Astrophysics, University of Nevada, Las Vegas, NV 89154, USA}
\affiliation{Dept. of Physics and Astronomy, University of Kansas, Lawrence, KS 66045, USA}
\affiliation{Department of Physics and Astronomy, UCLA, Los Angeles, CA 90095, USA}
\affiliation{Centre for Cosmology, Particle Physics and Phenomenology - CP3, Universit{\'e} catholique de Louvain, Louvain-la-Neuve, Belgium}
\affiliation{Department of Physics, Mercer University, Macon, GA 31207-0001, USA}
\affiliation{Dept. of Astronomy, University of Wisconsin{\textendash}Madison, Madison, WI 53706, USA}
\affiliation{Dept. of Physics and Wisconsin IceCube Particle Astrophysics Center, University of Wisconsin{\textendash}Madison, Madison, WI 53706, USA}
\affiliation{Institute of Physics, University of Mainz, Staudinger Weg 7, D-55099 Mainz, Germany}
\affiliation{Department of Physics, Marquette University, Milwaukee, WI, 53201, USA}
\affiliation{Institut f{\"u}r Kernphysik, Westf{\"a}lische Wilhelms-Universit{\"a}t M{\"u}nster, D-48149 M{\"u}nster, Germany}
\affiliation{Bartol Research Institute and Dept. of Physics and Astronomy, University of Delaware, Newark, DE 19716, USA}
\affiliation{Dept. of Physics, Yale University, New Haven, CT 06520, USA}
\affiliation{Columbia Astrophysics and Nevis Laboratories, Columbia University, New York, NY 10027, USA}
\affiliation{Dept. of Physics, University of Oxford, Parks Road, Oxford OX1 3PU, UK}
\affiliation{Dipartimento di Fisica e Astronomia Galileo Galilei, Universit{\`a} Degli Studi di Padova, 35122 Padova PD, Italy}
\affiliation{Dept. of Physics, Drexel University, 3141 Chestnut Street, Philadelphia, PA 19104, USA}
\affiliation{Physics Department, South Dakota School of Mines and Technology, Rapid City, SD 57701, USA}
\affiliation{Dept. of Physics, University of Wisconsin, River Falls, WI 54022, USA}
\affiliation{Dept. of Physics and Astronomy, University of Rochester, Rochester, NY 14627, USA}
\affiliation{Department of Physics and Astronomy, University of Utah, Salt Lake City, UT 84112, USA}
\affiliation{Oskar Klein Centre and Dept. of Physics, Stockholm University, SE-10691 Stockholm, Sweden}
\affiliation{Dept. of Physics and Astronomy, Stony Brook University, Stony Brook, NY 11794-3800, USA}
\affiliation{Dept. of Physics, Sungkyunkwan University, Suwon 16419, Korea}
\affiliation{Institute of Physics, Academia Sinica, Taipei, 11529, Taiwan}
\affiliation{Dept. of Physics and Astronomy, University of Alabama, Tuscaloosa, AL 35487, USA}
\affiliation{Dept. of Astronomy and Astrophysics, Pennsylvania State University, University Park, PA 16802, USA}
\affiliation{Dept. of Physics, Pennsylvania State University, University Park, PA 16802, USA}
\affiliation{Dept. of Physics and Astronomy, Uppsala University, Box 516, S-75120 Uppsala, Sweden}
\affiliation{Dept. of Physics, University of Wuppertal, D-42119 Wuppertal, Germany}
\affiliation{Deutsches Elektronen-Synchrotron DESY, Platanenallee 6, 15738 Zeuthen, Germany }

\author[0000-0001-6141-4205]{R. Abbasi}
\affiliation{Department of Physics, Loyola University Chicago, Chicago, IL 60660, USA}

\author[0000-0001-8952-588X]{M. Ackermann}
\affiliation{Deutsches Elektronen-Synchrotron DESY, Platanenallee 6, 15738 Zeuthen, Germany }

\author{J. Adams}
\affiliation{Dept. of Physics and Astronomy, University of Canterbury, Private Bag 4800, Christchurch, New Zealand}

\author[0000-0002-9714-8866]{S. K. Agarwalla}
\altaffiliation{also at Institute of Physics, Sachivalaya Marg, Sainik School Post, Bhubaneswar 751005, India}
\affiliation{Dept. of Physics and Wisconsin IceCube Particle Astrophysics Center, University of Wisconsin{\textendash}Madison, Madison, WI 53706, USA}

\author[0000-0003-2252-9514]{J. A. Aguilar}
\affiliation{Universit{\'e} Libre de Bruxelles, Science Faculty CP230, B-1050 Brussels, Belgium}

\author[0000-0003-0709-5631]{M. Ahlers}
\affiliation{Niels Bohr Institute, University of Copenhagen, DK-2100 Copenhagen, Denmark}

\author[0000-0002-9534-9189]{J.M. Alameddine}
\affiliation{Dept. of Physics, TU Dortmund University, D-44221 Dortmund, Germany}

\author{N. M. Amin}
\affiliation{Bartol Research Institute and Dept. of Physics and Astronomy, University of Delaware, Newark, DE 19716, USA}

\author{K. Andeen}
\affiliation{Department of Physics, Marquette University, Milwaukee, WI, 53201, USA}

\author[0000-0003-2039-4724]{G. Anton}
\affiliation{Erlangen Centre for Astroparticle Physics, Friedrich-Alexander-Universit{\"a}t Erlangen-N{\"u}rnberg, D-91058 Erlangen, Germany}

% ...

\date{\today}

\collaboration{393}{IceCube Collaboration}

%% file: IIn_catalogue.tex
SN1999bw                       &  2.70 &  0.79 & 1999-00-20 & 0.0032 &  9.80 & 1, 2\\
SN2002bu                       &  3.22 &  0.80 & 2002-00-28 & 0.0030 &  8.90 & 1, 2, 3\\
SN2008S                        &  5.39 &  1.05 & 2008-00-01 & 0.0012 &  5.60 & 4\\
SN2009kr                       &  1.36 & -0.27 & 2009-00-06 & 0.0075 & 16.00 & 5\\
SN2010jl                       &  2.54 &  0.17 & 2010-00-03 & 0.0117 & 49.00 & 6\\
SN2011an                       &  2.09 &  0.29 & 2011-00-01 & 0.0170 & 73.00 & 7\\
SN2011ht                       &  2.65 &  0.90 & 2011-00-29 & 0.0046 & 19.20 & 8\\
SN2012ab                       &  3.24 &  0.10 & 2012-00-31 & 0.0190 & 81.00 & 9\\
SN2013by                       &  4.29 & -1.05 & 2013-00-23 & 0.0038 & 14.80 & 10, 11\\
SN2013gc                       &  2.13 & -0.49 & 2013-00-07 & 0.0044 & 15.10 & 12\\
PSN J14041297-0938168          &  3.68 & -0.17 & 2013-00-20 & 0.0038 & 12.55 & 13\\
CSS140111:060437-123740        &  1.59 & -0.22 & 2013-00-24 & 0.0084 & 32.88 & 13\\
SN2014G                        &  2.86 &  0.95 & 2014-00-14 & 0.0045 & 20.00 & 14\\
MASTER OT J044212.20+230616.7  &  1.23 &  0.40 & 2014-00-21 & 0.0170 & 72.00 & 15\\
SN2015da                       &  3.63 &  0.69 & 2015-00-09 & 0.0079 & 32.14 & 16, 17\\

%% file: IIn_reference_map.tex
(1) \citet{kochanekUnmaskingSupernovaImpostors2012}, (2) \citet{smithLuminousBlueVariable2011a}, (3) \citet{szczygielSN2002buAnother2012}, (4) \citet{stanishevSupernova2008SNGC2008}, (5) \citet{steeleSupernovae2009kn2009ko2009}, (6) \citet{benettiSupernova2010jlUGC2010}, (7) \citet{marionSupernova2011anUGC2011}, (8) \citet{prietoSupernova2011htUGC2011}, (9) \citet{bilinskiSN2012abPeculiarType2018}, (10) \citet{marguttiXrayEmissionPosition2013}, (11) \citet{parkerSupernova2013byESO2013}, (12) \citet{antezanaSupernova2013gcESO2013}, (13) \citet{challisSpectraIDRecent2013a}, (14) \citet{denisenkoSupernova2014GNGC2014}, (15) \citet{shivversSpectroscopicClassificationMASTER2014}, (16) \citet{zhangSpectroscopicClassificationPSN2015}, (17) \citet{tartagliaLonglivedTypeIIn2020}

%% file: IIP_catalogue.tex
SN1999em                       &  1.23 & -0.05 & 1999-00-29 & 0.0034 &  7.50 & 1\\
SN2004dj                       &  2.00 &  1.14 & 2004-00-31 & 0.0014 &  3.50 & 2\\
SN2004et                       &  5.39 &  1.05 & 2004-00-27 & 0.0022 &  7.70 & 3, 4\\
SN2005cs                       &  3.53 &  0.82 & 2005-00-28 & 0.0030 &  7.10 & 5, 6\\
SN2006ov                       &  3.24 &  0.08 & 2006-00-24 & 0.0062 & 14.00 & 7\\
SN2008bk                       &  6.27 & -0.57 & 2008-00-25 & 0.0018 &  4.00 & 8\\
SN2009js                       &  0.64 &  0.32 & 2009-00-11 & 0.0060 & 16.00 & 9\\
SN2009md                       &  2.83 &  0.22 & 2009-00-05 & 0.0046 & 18.00 & 10\\
SN2009mf                       &  0.27 &  0.83 & 2009-00-07 & 0.0087 & 23.00 & 11\\
SN2011dq                       &  0.26 & -0.13 & 2011-00-15 & 0.0055 & 24.40 & 12\\
SN2012A                        &  2.73 &  0.30 & 2012-00-07 & 0.0034 &  9.80 & 13\\
SN2012aw                       &  2.81 &  0.20 & 2012-00-16 & 0.0036 &  9.90 & 14\\
SNhunt141                      &  3.57 & -0.31 & 2012-00-24 & 0.0040 & 18.00 & 15\\
SN2012ec                       &  0.72 & -0.13 & 2012-00-12 & 0.0057 & 18.76 & 16\\
SN2013ab                       &  3.81 &  0.17 & 2013-00-17 & 0.0063 & 23.64 & 17\\
SN2013am                       &  2.96 &  0.23 & 2013-00-21 & 0.0037 & 12.77 & 18\\
SN2013bu                       &  5.92 &  0.60 & 2013-00-21 & 0.0027 & 12.07 & 19\\
SN2013ej                       &  0.42 &  0.28 & 2013-00-25 & 0.0020 &  9.00 & 20\\
SN2011ja                       &  3.43 & -0.86 & 2014-00-14 & 0.0018 &  3.36 & 21\\
SN2014bc                       &  3.22 &  0.83 & 2014-00-19 & 0.0025 &  7.60 & 22\\

%% file: IIP_reference_map.tex
(1) \citet{jhaSupernova1999emNGC1999}, (2) \citet{patatSupernova2004djNGC2004}, (3) \citet{zwitterSupernova2004etNGC2004}, (4) \citet{liProgenitorTypeII2005}, (5) \citet{modjazSupernova2005csM512005}, (6) \citet{pastorelloSN2005csM512009}, (7) \citet{liProgenitorsTwoType2007}, (8) \citet{morrellSupernovae2008bk2008br2008}, (9) \citet{gandhiSN2009jsCrossroads2013}, (10) \citet{sollermanSupernova2009mdNGC2009}, (11) \citet{steeleSupernovae2009lu2009md2009}, (12) \citet{valentiSupernova2011dqNGC2011}, (13) \citet{stanishevSupernova2012ANGC2012}, (14) \citet{quadriSupernova2012awM952012}, (15) \citet{cellier-holzemPESSTOSpectroscopicClassification2012}, (16) \citet{monardSupernova2012ecNGC2012}, (17) \citet{boseSN2013abNormal2015}, (18) \citet{benettiPSNJ1118569513034942013}, (19) \citet{itagakiSupernova2013buNGC2013}, (20) \citet{dhunganaExtensiveSpectroscopyPhotometry2016}, (21) \citet{andrewsEarlyDustFormation2016}, (22) \citet{ochnerAsiagoClassificationPS114xz2014}

%% file: Ibc_catalogue.tex
SN2007gr                       &  0.71 &  0.65 & 2007-00-15 & 0.0027 &  9.30 & 1, 2\\
SN2008ax                       &  3.28 &  0.73 & 2008-00-03 & 0.0029 &  9.60 & 3, 4\\
SN2008dv                       &  0.95 &  1.27 & 2008-00-01 & 0.0084 &  4.20 & 5\\
SN2009dq                       &  2.66 & -1.17 & 2009-00-24 & 0.0046 & 16.00 & 6\\
SN2009gj                       &  0.13 & -0.58 & 2009-00-21 & 0.0053 & 17.00 & 7\\
SN2009mk                       &  0.03 & -0.72 & 2009-00-15 & 0.0050 & 22.00 & 8, 9\\
SN2009mu                       &  2.58 & -0.58 & 2009-00-21 & 0.0098 & 25.00 & 10\\
SN2010br                       &  3.16 &  0.78 & 2010-00-10 & 0.0033 & 13.00 & 11\\
SN2010gi                       &  4.55 &  1.32 & 2010-00-18 & 0.0041 & 18.20 & 12\\
SN2011dh                       &  3.53 &  0.82 & 2011-00-01 & 0.0025 &  8.40 & 5\\
SN2011jm                       &  3.38 &  0.05 & 2011-00-24 & 0.0041 & 14.00 & 13\\
SN2012P                        &  3.93 &  0.03 & 2012-00-22 & 0.0055 & 20.10 & 14, 15\\
SN2012cw                       &  2.68 &  0.06 & 2012-00-14 & 0.0055 & 19.92 & 16, 17\\
SN2012fh                       &  2.81 &  0.43 & 2012-00-18 & 0.0029 &  8.58 & 18, 19, 20\\
SN2013df                       &  3.26 &  0.55 & 2013-00-07 & 0.0033 & 10.58 & 21, 22\\
iPTF13bvn                      &  3.93 &  0.03 & 2013-00-17 & 0.0055 & 19.94 & 15, 23, 24, 25\\
MASTER OT J120451.50+265946.6  &  3.16 &  0.47 & 2013-00-02 & 0.0029 &  8.38 & 26, 27, 28\\
SN2013ge                       &  2.77 &  0.38 & 2013-00-08 & 0.0054 & 19.34 & 29, 30\\
SN2014C                        &  5.92 &  0.60 & 2014-00-05 & 0.0037 & 12.07 & 31, 32, 33\\

%% file: Ibc_reference_map.tex
(1) \citet{chornockSupernova2007grNGC2007}, (2) \citet{valentiCarbonrichTypeIc2008}, (3) \citet{chornockSupernova2008axNGC2008}, (4) \citet{pastorelloTypeIIbSN2008}, (5) \citet{silvermanSupernovae2008dv2008dz2008}, (6) \citet{andersonSupernovae2009dp2009dq2009}, (7) \citet{stockdaleSupernova2009gjNGC2009}, (8) \citet{chornockSupernova2009mkPgc2009}, (9) \citet{marplesSupernova2009mkPgc2009}, (10) \citet{stritzingerSupernova2009muESO2010}, (11) \citet{maxwellSupernova2010brNGC2010}, (12) \citet{yamanakaSupernova2010giIC2010}, (13) \citet{foleySupernova2011jmNGC2011}, (14) \citet{borsatoSupernova2012PNGC2012}, (15) \citet{fremlingPTF12osIPTF13bvnTwo2016}, (16) \citet{itagakiSupernova2012cwNGC2012}, (17) \citet{wangSupernova2012cwNGC2012}, (18) \citet{johnsonProgenitorTypeIbc2017}, (19) \citet{takakiSupernova2012fhNGC2012}, (20) \citet{tomasellaSupernova2012fhNGC2012}, (21) \citet{ciabattariSupernova2013dfNGC2013}, (22) \citet{vandykTypeIIbSupernova2014}, (23) \citet{caoDiscoveryProgenitorEarly2013}, (24) \citet{milisavljevicOpticalSpectroscopyIPTF13bvn2013}, (25) \citet{srivastavOpticalObservationsFast2014}, (26) \citet{chandraTypeIbSupernova2019}, (27) \citet{singhObservationalPropertiesType2019}, (28) \citet{srivastavSpectroscopicClassificationMASTER2014}, (29) \citet{droutDoublepeakedSN2013ge2016}, (30) \citet{nakanoSupernova2013geNGC2013}, (31) \citet{kimSupernova2014CNGC2014}, (32) \citet{milisavljevicMetamorphosisSN2014C2015}, (33) \citet{tinyanontSystematicStudyMidinfrared2016}

%% file: p-values.tex
IIn &   -  &  8.6 &  6.3 &  $>$50 &  $>$50 &  $>$50 & 30.1 \\ 
IIP &   -  & 48.6 &  $>$50 & 27.6 &  $>$50 &  $>$50 & 21.6 \\ 
Stripped-envelope &  $>$50 &  $>$50 &  $>$50 & 34.8 &   -  &   -  &   -  \\ 

%% file: CCSN2022_main.bbl
\begin{thebibliography}{}
\expandafter\ifx\csname natexlab\endcsname\relax\def\natexlab#1{#1}\fi
\providecommand{\url}[1]{\href{#1}{#1}}
\providecommand{\dodoi}[1]{doi:~\href{http://doi.org/#1}{\nolinkurl{#1}}}
\providecommand{\doeprint}[1]{\href{http://ascl.net/#1}{\nolinkurl{http://ascl.net/#1}}}
\providecommand{\doarXiv}[1]{\href{https://arxiv.org/abs/#1}{\nolinkurl{https://arxiv.org/abs/#1}}}

\bibitem[{Aartsen {et~al.}(2013)}]{Aartsen:2013jdh}
Aartsen, M.~G., {et~al.} 2013, Science, 342, 1242856,
  \dodoi{10.1126/science.1242856}

\bibitem[{Aartsen {et~al.}(2015)}]{Aartsen:2015knd}
---. 2015, Astrophys. J., 809, 98, \dodoi{10.1088/0004-637X/809/1/98}

\bibitem[{{Aartsen} {et~al.}(2017{\natexlab{a}}){Aartsen}, {Abraham},
  {Ackermann}, {Adams}, {Aguilar}, {Ahlers}, {Ahrens}, {Altmann}, {Andeen},
  {Anderson}, {Ansseau}, {Anton}, {Archinger}, {Arguelles}, {Arlen},
  {Auffenberg}, {Axani}, {Bai}, {Barwick}, {Baum}, {Bay}, {Beatty}, {Becker
  Tjus}, {Becker}, {BenZvi}, {Berghaus}, {Berley}, {Bernardini}, {Bernhard},
  {Besson}, {Binder}, {Bindig}, {Bissok}, {Blaufuss}, {Blot}, {Boersma},
  {Bohm}, {B{\"o}rner}, {Bos}, {Bose}, {B{\"o}ser}, {Botner}, {Braun},
  {Brayeur}, {Bretz}, {Burgman}, {Casey}, {Casier}, {Cheung}, {Chirkin},
  {Christov}, {Clark}, {Classen}, {Coenders}, {Collin}, {Conrad}, {Cowen},
  {Cruz Silva}, {Daughhetee}, {Davis}, {Day}, {de Andr{\'e}}, {De Clercq}, {del
  Pino Rosendo}, {Dembinski}, {De Ridder}, {Desiati}, {de Vries}, {de
  Wasseige}, {de With}, {DeYoung}, {D{\'\i}az-V{\'e}lez}, {di Lorenzo},
  {Dujmovic}, {Dumm}, {Dunkman}, {Eberhardt}, {Ehrhardt}, {Eichmann}, {Euler},
  {Evenson}, {Fahey}, {Fazely}, {Feintzeig}, {Felde}, {Filimonov}, {Finley},
  {Flis}, {F{\"o}sig}, {Franckowiak}, {Fuchs}, {Gaisser}, {Gaior}, {Gallagher},
  {Gerhardt}, {Ghorbani}, {Giang}, {Gladstone}, {Glagla}, {Gl{\"u}senkamp},
  {Goldschmidt}, {Golup}, {Gonzalez}, {G{\'o}ra}, {Grant}, {Griffith}, {Haack},
  {Haj Ismail}, {Hallgren}, {Halzen}, {Hansen}, {Hansmann}, {Hansmann},
  {Hanson}, {Hebecker}, {Heereman}, {Helbing}, {Hellauer}, {Hickford},
  {Hignight}, {Hill}, {Hoffman}, {Hoffmann}, {Holzapfel}, {Homeier}, {Hoshina},
  {Huang}, {Huber}, {Huelsnitz}, {Hultqvist}, {In}, {Ishihara}, {Jacobi},
  {Japaridze}, {Jeong}, {Jero}, {Jones}, {Jurkovic}, {Kappes}, {Karg}, {Karle},
  {Katz}, {Kauer}, {Keivani}, {Kelley}, {Kemp}, {Kheirandish}, {Kim},
  {Kintscher}, {Kiryluk}, {Kittler}, {Klein}, {Kohnen}, {Koirala}, {Kolanoski},
  {Konietz}, {K{\"o}pke}, {Kopper}, {Kopper}, {Koskinen}, {Kowalski}, {Krings},
  {Kroll}, {Kr{\"u}ckl}, {Kr{\"u}ger}, {Kunnen}, {Kunwar}, {Kurahashi},
  {Kuwabara}, {Labare}, {Lanfranchi}, {Larson}, {Lennarz}, {Lesiak-Bzdak},
  {Leuermann}, {Leuner}, {Lu}, {L{\"u}nemann}, {Madsen}, {Maggi}, {Mahn},
  {Mancina}, {Mandelartz}, {Maruyama}, {Mase}, {Maunu}, {McNally}, {Meagher},
  {Medici}, {Meier}, {Meli}, {Menne}, {Merino}, {Meures}, {Miarecki},
  {Middell}, {Mohrmann}, {Montaruli}, {Moulai}, {Nahnhauer}, {Naumann}, {Neer},
  {Niederhausen}, {Nowicki}, {Nygren}, {Obertacke Pollmann}, {Olivas},
  {Omairat}, {O'Murchadha}, {Palczewski}, {Pandya}, {Pankova}, {Penek},
  {Pepper}, {P{\'e}rez de los Heros}, {Pfendner}, {Pieloth}, {Pinat},
  {Posselt}, {Price}, {Przybylski}, {Quinnan}, {Raab}, {R{\"a}del}, {Rameez},
  {Rawlins}, {Reimann}, {Relich}, {Resconi}, {Rhode}, {Richman}, {Riedel},
  {Robertson}, {Rongen}, {Rott}, {Ruhe}, {Ryckbosch}, {Rysewyk}, {Sabbatini},
  {Sanchez Herrera}, {Sandrock}, {Sandroos}, {Sarkar}, {Satalecka}, {Schimp},
  {Schlunder}, {Schmidt}, {Schoenen}, {Sch{\"o}neberg}, {Sch{\"o}nwald},
  {Schumacher}, {Seckel}, {Seunarine}, {Soldin}, {Song}, {Spiczak}, {Spiering},
  {Stahlberg}, {Stamatikos}, {Stanev}, {Stasik}, {Steuer}, {Stezelberger},
  {Stokstad}, {St{\"o}{\ss}l}, {Str{\"o}m}, {Strotjohann}, {Sullivan},
  {Sutherland}, {Taavola}, {Taboada}, {Tatar}, {Ter-Antonyan}, {Terliuk},
  {Te{\v{s}}i{\'c}}, {Tilav}, {Toale}, {Tobin}, {Toscano}, {Tosi},
  {Tselengidou}, {Turcati}, {Unger}, {Usner}, {Vallecorsa}, {Vandenbroucke},
  {van Eijndhoven}, {Vanheule}, {van Rossem}, {van Santen}, {Veenkamp},
  {Vehring}, {Voge}, {Vraeghe}, {Walck}, {Wallace}, {Wallraff}, {Wandkowsky},
  {Weaver}, {Wendt}, {Westerhoff}, {Whelan}, {Wickmann}, {Wiebe}, {Wiebusch},
  {Wille}, {Williams}, {Wills}, {Wissing}, {Wolf}, {Wood}, {Woolsey},
  {Woschnagg}, {Xu}, {Xu}, {Xu}, {Yanez}, {Yodh}, {Yoshida}, {Zoll}, \&
  {IceCube Collaboration}}]{2017ApJ...835...45A}
{Aartsen}, M.~G., {Abraham}, K., {Ackermann}, M., {et~al.} 2017{\natexlab{a}},
  \apj, 835, 45, \dodoi{10.3847/1538-4357/835/1/45}

\bibitem[{{Aartsen} {et~al.}(2017{\natexlab{b}}){Aartsen}, {Ackermann},
  {Adams}, {Aguilar}, {Ahlers}, {Ahrens}, {Samarai}, {Altmann}, {Andeen},
  {Anderson}, \& et~al.}]{2017ApJ...843..112A}
{Aartsen}, M.~G., {Ackermann}, M., {Adams}, J., {et~al.} 2017{\natexlab{b}},
  \apj, 843, 112, \dodoi{10.3847/1538-4357/aa7569}

\bibitem[{Aartsen {et~al.}(2017{\natexlab{a}})}]{Aartsen:2016nxy}
Aartsen, M.~G., {et~al.} 2017{\natexlab{a}}, JINST, 12, P03012,
  \dodoi{10.1088/1748-0221/12/03/P03012}

\bibitem[{Aartsen {et~al.}(2017{\natexlab{b}})}]{Aartsen:2016oji}
---. 2017{\natexlab{b}}, Astrophys. J., 835, 151,
  \dodoi{10.3847/1538-4357/835/2/151}

\bibitem[{Aartsen {et~al.}(2018{\natexlab{a}})}]{MWScience}
---. 2018{\natexlab{a}}, Science, 361, X

\bibitem[{Aartsen {et~al.}(2018{\natexlab{b}})}]{ICScience}
---. 2018{\natexlab{b}}, Science, 361, 147

\bibitem[{Ade {et~al.}(2016)}]{Ade:2015xua}
Ade, P. A.~R., {et~al.} 2016, Astron. Astrophys., 594, A13,
  \dodoi{10.1051/0004-6361/201525830}

\bibitem[{Ahlers \& Halzen(2014)}]{Ahlers:2014ioa}
Ahlers, M., \& Halzen, F. 2014, Phys. Rev., D90, 043005,
  \dodoi{10.1103/PhysRevD.90.043005}

\bibitem[{Anderson {et~al.}(2009)Anderson, Morrell, Folatelli, \&
  Stritzinger}]{andersonSupernovae2009dp2009dq2009}
Anderson, J., Morrell, N., Folatelli, G., \& Stritzinger, M. 2009, Central
  Bureau Electronic Telegrams, 1789, 1

\bibitem[{{Ando} \& {Beacom}(2005)}]{2005PhRvL..95f1103A}
{Ando}, S., \& {Beacom}, J.~F. 2005, Physical Review Letters, 95, 061103,
  \dodoi{10.1103/PhysRevLett.95.061103}

\bibitem[{Andrews {et~al.}(2016)Andrews, Krafton, Clayton, Montiel, Wesson,
  Sugerman, Barlow, Matsuura, \& Drass}]{andrewsEarlyDustFormation2016}
Andrews, J.~E., Krafton, K.~M., Clayton, G.~C., {et~al.} 2016, Monthly Notices
  of the Royal Astronomical Society, 457, 3241, \dodoi{10.1093/mnras/stw164}

\bibitem[{Antezana {et~al.}(2013)Antezana, Hamuy, Gonzalez, Cartier, Forster,
  Carrasco, Pignata, Apostolovski, Paillas, Varela, Bufano, Olivares, Takats,
  Aros, Conuel, Folatelli, Reichart, Haislip, Moore, LaCluyze, \&
  Morrell}]{antezanaSupernova2013gcESO2013}
Antezana, R., Hamuy, M., Gonzalez, L., {et~al.} 2013, Central Bureau Electronic
  Telegrams, 3699, 1

\bibitem[{Benetti {et~al.}(2010)Benetti, Bufano, Vinko, Marion, Pritchard,
  Wheeler, Chatzopoulos, \& Shetrone}]{benettiSupernova2010jlUGC2010}
Benetti, S., Bufano, F., Vinko, J., {et~al.} 2010, Central Bureau Electronic
  Telegrams, 2536, 1

\bibitem[{Benetti {et~al.}(2013)Benetti, Tomasella, Pastorello, Cappellaro,
  Turatto, \& Ochner}]{benettiPSNJ1118569513034942013}
Benetti, S., Tomasella, L., Pastorello, A., {et~al.} 2013, The Astronomer's
  Telegram, 4909, 1

\bibitem[{Bilinski {et~al.}(2018)Bilinski, Smith, Williams, Smith, Zheng,
  Graham, Mauerhan, Andrews, Filippenko, Akerlof, Chatzopoulos, Hoffman, Huk,
  Leonard, Marion, Milne, Quimby, Silverman, Vink{\'o}, Wheeler, \&
  Yuan}]{bilinskiSN2012abPeculiarType2018}
Bilinski, C., Smith, N., Williams, G.~G., {et~al.} 2018, Monthly Notices of the
  Royal Astronomical Society, 475, 1104, \dodoi{10.1093/mnras/stx3214}

\bibitem[{Borsato {et~al.}(2012)Borsato, Nascimbeni, Benetti, Pastorello,
  Valenti, Tomasella, Cappellaro, Ochner, \&
  Turatto}]{borsatoSupernova2012PNGC2012}
Borsato, L., Nascimbeni, V., Benetti, S., {et~al.} 2012, Central Bureau
  Electronic Telegrams, 2993, 2

\bibitem[{Bose {et~al.}(2015)Bose, Valenti, Misra, Pumo, Zampieri, Sand, Kumar,
  Pastorello, Sutaria, Maccarone, Kumar, Graham, Howell, Ochner, Chandola, \&
  Pandey}]{boseSN2013abNormal2015}
Bose, S., Valenti, S., Misra, K., {et~al.} 2015, Monthly Notices of the Royal
  Astronomical Society, 450, 2373, \dodoi{10.1093/mnras/stv759}

\bibitem[{{Braun} {et~al.}(2010){Braun}, {Baker}, {Dumm}, {Finley}, {Karle}, \&
  {Montaruli}}]{Braun:2009aa}
{Braun}, J., {Baker}, M., {Dumm}, J., {et~al.} 2010, Astroparticle Physics, 33,
  175, \dodoi{10.1016/j.astropartphys.2010.01.005}

\bibitem[{Braun {et~al.}(2008)Braun, Dumm, De~Palma, Finley, Karle, \&
  Montaruli}]{Braun:2008bg}
Braun, J., Dumm, J., De~Palma, F., {et~al.} 2008, Astropart. Phys., 29, 299,
  \dodoi{10.1016/j.astropartphys.2008.02.007}

\bibitem[{Cao {et~al.}(2013)Cao, Kasliwal, Arcavi, Horesh, Hancock, Valenti,
  Cenko, Kulkarni, {Gal-Yam}, Gorbikov, Ofek, Sand, Yaron, Graham, Silverman,
  Wheeler, Marion, Walker, Mazzali, Howell, Li, Kong, Bloom, Nugent, Surace,
  Masci, Carpenter, Degenaar, \& Gelino}]{caoDiscoveryProgenitorEarly2013}
Cao, Y., Kasliwal, M.~M., Arcavi, I., {et~al.} 2013, The Astrophysical Journal,
  775, L7, \dodoi{10.1088/2041-8205/775/1/L7}

\bibitem[{{Cellier-Holzem} {et~al.}(2012){Cellier-Holzem}, Smartt, Inserra,
  Fraser, Wright, Young, Smith, Pastorello, Valenti, Benetti, Taubenberger,
  Sullivan, {Gal-Yam}, Yaron, Howerton, Baltay, Ellman, Hadjiyska, McKinnon,
  Rabinowitz, Feindt, Kowalski, \&
  Nugent}]{cellier-holzemPESSTOSpectroscopicClassification2012}
{Cellier-Holzem}, F., Smartt, S.~J., Inserra, C., {et~al.} 2012, The
  Astronomer's Telegram, 4300, 1

\bibitem[{Challis(2013)}]{challisSpectraIDRecent2013a}
Challis, P. 2013, The Astronomer's Telegram, 5700, 1

\bibitem[{Chandra {et~al.}(2019)Chandra, Nayana, Bj{\"o}rnsson, Taddia,
  Lundqvist, Ray, \& Shappee}]{chandraTypeIbSupernova2019}
Chandra, P., Nayana, A.~J., Bj{\"o}rnsson, C.~I., {et~al.} 2019, The
  Astrophysical Journal, 877, 79, \dodoi{10.3847/1538-4357/ab1900}

\bibitem[{Chornock \& Berger(2009)}]{chornockSupernova2009mkPgc2009}
Chornock, R., \& Berger, E. 2009, Central Bureau Electronic Telegrams, 2086, 1

\bibitem[{Chornock {et~al.}(2007)Chornock, Filippenko, Li, Foley, Reitzel, \&
  Rich}]{chornockSupernova2007grNGC2007}
Chornock, R., Filippenko, A.~V., Li, W., {et~al.} 2007, Central Bureau
  Electronic Telegrams, 1036, 1

\bibitem[{Chornock {et~al.}(2008)Chornock, Filippenko, Li, Foley, Stockton,
  Moran, Hodge, \& Merriman}]{chornockSupernova2008axNGC2008}
---. 2008, Central Bureau Electronic Telegrams, 1298, 1

\bibitem[{Ciabattari {et~al.}(2013)Ciabattari, Mazzoni, Donati, Petroni,
  Foglia, Galli, Cenko, Clubb, Zheng, Kelly, Filippenko, \&
  Van~Dyk}]{ciabattariSupernova2013dfNGC2013}
Ciabattari, F., Mazzoni, E., Donati, S., {et~al.} 2013, Central Bureau
  Electronic Telegrams, 3557, 1

\bibitem[{Coenders(2016)}]{Coenders2016}
Coenders, S. 2016, Dissertation, Technische Universit{\"a}t M{\"u}nchen,
  M{\"u}nchen

\bibitem[{Denisenko {et~al.}(2014)Denisenko, Lipunov, Gorbovskoy, Lake, Yusa,
  Kadota, Itoh, Moritani, Kawabata, Yamanaka, Ochner, Siviero, Tomasella,
  Benetti, Cappellaro, {Elias-Rosa}, Pastorello, Tartaglia, Terreran, Turatto,
  \& Nakano}]{denisenkoSupernova2014GNGC2014}
Denisenko, D., Lipunov, V., Gorbovskoy, E., {et~al.} 2014, Central Bureau
  Electronic Telegrams, 3787, 2

\bibitem[{{Denton} \& {Tamborra}(2018)}]{2018ApJ...855...37D}
{Denton}, P.~B., \& {Tamborra}, I. 2018, \apj, 855, 37,
  \dodoi{10.3847/1538-4357/aaab4a}

\bibitem[{Dhungana {et~al.}(2016)Dhungana, Kehoe, Vinko, Silverman, Wheeler,
  Zheng, Marion, Fox, Akerlof, Biro, Borkovits, Cenko, Clubb, Filippenko,
  Ferrante, Gibson, Graham, Hegedus, Kelly, Kelemen, Lee, Marschalko,
  Moln{\'a}r, Nagy, Ordasi, Pal, Sarneczky, Shivvers, Szakats, Szalai,
  {Szegedi-Elek}, Sz{\'e}kely, Szing, Tak{\'a}ts, \&
  Vida}]{dhunganaExtensiveSpectroscopyPhotometry2016}
Dhungana, G., Kehoe, R., Vinko, J., {et~al.} 2016, The Astrophysical Journal,
  822, 6, \dodoi{10.3847/0004-637X/822/1/6}

\bibitem[{Drout {et~al.}(2016)Drout, Milisavljevic, Parrent, Margutti, Kamble,
  Soderberg, Challis, Chornock, Fong, Frank, Gehrels, Graham, Hsiao, Itagaki,
  Kasliwal, Kirshner, Macomb, Marion, Norris, \&
  Phillips}]{droutDoublepeakedSN2013ge2016}
Drout, M.~R., Milisavljevic, D., Parrent, J., {et~al.} 2016, The Astrophysical
  Journal, 821, 57, \dodoi{10.3847/0004-637X/821/1/57}

\bibitem[{{Faran} {et~al.}(2014){Faran}, {Poznanski}, {Filippenko}, {Chornock},
  {Foley}, {Ganeshalingam}, {Leonard}, {Li}, {Modjaz}, {Nakar}, {Serduke}, \&
  {Silverman}}]{faran2014}
{Faran}, T., {Poznanski}, D., {Filippenko}, A.~V., {et~al.} 2014, \mnras, 442,
  844, \dodoi{10.1093/mnras/stu955}

\bibitem[{Foley \& Fong(2011)}]{foleySupernova2011jmNGC2011}
Foley, R.~J., \& Fong, W. 2011, Central Bureau Electronic Telegrams, 2962, 2

\bibitem[{Fremling {et~al.}(2016)Fremling, Sollerman, Taddia, Ergon, Fraser,
  Karamehmetoglu, Valenti, Jerkstrand, Arcavi, Bufano, Elias~Rosa, Filippenko,
  Fox, {Gal-Yam}, Howell, Kotak, Mazzali, Milisavljevic, Nugent, Nyholm, Pian,
  \& Smartt}]{fremlingPTF12osIPTF13bvnTwo2016}
Fremling, C., Sollerman, J., Taddia, F., {et~al.} 2016, Astronomy and
  Astrophysics, 593, A68, \dodoi{10.1051/0004-6361/201628275}

\bibitem[{Gandhi {et~al.}(2013)Gandhi, Yamanaka, Tanaka, Nozawa, Kawabata,
  Saviane, Maeda, Moriya, Hattori, Sasada, \&
  Itoh}]{gandhiSN2009jsCrossroads2013}
Gandhi, P., Yamanaka, M., Tanaka, M., {et~al.} 2013, The Astrophysical Journal,
  767, 166, \dodoi{10.1088/0004-637X/767/2/166}

\bibitem[{{Graham} {et~al.}(2019){Graham}, {Kulkarni}, {Bellm}, {Adams},
  {Barbarino}, {Blagorodnova}, {Bodewits}, {Bolin}, {Brady}, {Cenko}, {Chang},
  {Coughlin}, {De}, {Eadie}, {Farnham}, {Feindt}, {Franckowiak}, {Fremling},
  {Gal-yam}, {Gezari}, {Ghosh}, {Goldstein}, {Golkhou}, {Goobar}, {Ho},
  {Huppenkothen}, {Ivezic}, {Jones}, {Juric}, {Kaplan}, {Kasliwal}, {Kelley},
  {Kupfer}, {Lee}, {Lin}, {Lunnan}, {Mahabal}, {Miller}, {Ngeow}, {Nugent},
  {Ofek}, {Prince}, {Rauch}, {van Roestel}, {Schulze}, {Singer}, {Sollerman},
  {Taddia}, {Yan}, {Ye}, {Yu}, {Andreoni}, {Barlow}, {Bauer}, {Beck},
  {Belicki}, {Biswas}, {Brinnel}, {Brooke}, {Bue}, {Bulla}, {Burdge},
  {Burruss}, {Connolly}, {Cromer}, {Cunningham}, {Dekany}, {Delacroix},
  {Desai}, {Duev}, {Hacopians}, {Hale}, {Helou}, {Henning}, {Hover},
  {Hillenbrand}, {Howell}, {Hung}, {Imel}, {Ip}, {Jackson}, {Kaspi}, {Kaye},
  {Kowalski}, {Kramer}, {Kuhn}, {Land ry}, {Laher}, {Mao}, {Masci},
  {Monkewitz}, {Murphy}, {Nordin}, {Patterson}, {Penprase}, {Porter},
  {Rebbapragada}, {Reiley}, {Riddle}, {Rigault}, {Rodriguez}, {Rusholme}, {van
  Santen}, {Shupe}, {Smith}, {Soumagnac}, {Stein}, {Surace}, {Szkody}, {Terek},
  {van Sistine}, {van Velzen}, {Vestrand}, {Walters}, {Ward}, {Zhang}, \&
  {Zolkower}}]{2019arXiv190201945G}
{Graham}, M.~J., {Kulkarni}, S.~R., {Bellm}, E.~C., {et~al.} 2019, arXiv
  e-prints, arXiv:1902.01945.
\newblock \doarXiv{1902.01945}

\bibitem[{{Guillochon} {et~al.}(2017){Guillochon}, {Parrent}, {Kelley}, \&
  {Margutti}}]{Guillochon:2016rhj}
{Guillochon}, J., {Parrent}, J., {Kelley}, L.~Z., \& {Margutti}, R. 2017, \apj,
  835, 64, \dodoi{10.3847/1538-4357/835/1/64}

\bibitem[{{Hogg}(1999)}]{Hogg:1999ad}
{Hogg}, D.~W. 1999, arXiv e-prints, astro.
\newblock \doarXiv{astro-ph/9905116}

\bibitem[{{IceCube Collaboration} {et~al.}(2022){IceCube Collaboration},
  Abbasi, Ackermann, Adams, Aguilar, Ahlers, Ahrens, Alameddine, Alispach,
  Alves, Amin, Andeen, Anderson, Anton, Arg{\"u}elles, Ashida, Axani, Bai,
  Balagopal~V., Barbano, Barwick, Bastian, Basu, Baur, Bay, Beatty, Becker,
  Becker~Tjus, Bellenghi, BenZvi, Berley, Bernardini, Besson, Binder, Bindig,
  Blaufuss, Blot, Boddenberg, Bontempo, Borowka, B{\"o}ser, Botner,
  B{\"o}ttcher, Bourbeau, Bradascio, Braun, Brinson, Bron, {Brostean-Kaiser},
  Browne, Burgman, Burley, Busse, Campana, {Carnie-Bronca}, Chen, Chen,
  Chirkin, Choi, Clark, Clark, Classen, Coleman, Collin, Conrad, Coppin,
  Correa, Cowen, Cross, Dappen, Dave, De~Clercq, DeLaunay, Delgado~L{\'o}pez,
  Dembinski, Deoskar, Desai, Desiati, {de Vries}, {de Wasseige}, {de With},
  DeYoung, Diaz, {D{\'i}az-V{\'e}lez}, Dittmer, Dujmovic, Dunkman, DuVernois,
  Dvorak, Ehrhardt, Eller, Engel, Erpenbeck, Evans, Evenson, Fan, Fazely,
  Fedynitch, Feigl, Fiedlschuster, Fienberg, Filimonov, Finley, Fischer, Fox,
  Franckowiak, Friedman, Fritz, F{\"u}rst, Gaisser, Gallagher, Ganster, Garcia,
  Garrappa, Gerhardt, Ghadimi, Glaser, Glauch, Gl{\"u}senkamp, Goldschmidt,
  Gonzalez, Goswami, Grant, Gr{\'e}goire, Griswold, G{\"u}nther, Gutjahr,
  Haack, Hallgren, Halliday, Halve, Halzen, Ha~Minh, Hanson, Hardin, Harnisch,
  Haungs, Hebecker, Helbing, Henningsen, Hettinger, Hickford, Hignight, Hill,
  Hill, Hoffman, Hoffmann, {Hokanson-Fasig}, Hoshina, Huang, Huber, Huber,
  Hultqvist, H{\"u}nnefeld, Hussain, Hymon, In, Iovine, Ishihara, Jansson,
  Japaridze, Jeong, Jin, Jones, Kang, Kang, Kang, Kappes, Kappesser, Kardum,
  Karg, Karl, Karle, Katz, Kauer, Kellermann, Kelley, Kheirandish, Kin,
  Kintscher, Kiryluk, Klein, Koirala, Kolanoski, Kontrimas, K{\"o}pke, Kopper,
  Kopper, Koskinen, Koundal, Kovacevich, Kowalski, Kozynets, Kun, Kurahashi,
  Lad, Lagunas~Gualda, Lanfranchi, Larson, Lauber, Lazar, Lee, Leonard,
  Leszczy{\'n}ska, Li, Lincetto, Liu, Liubarska, Lohfink, Lozano~Mariscal, Lu,
  Lucarelli, Ludwig, Luszczak, Lyu, Ma, Madsen, Mahn, Makino, Mancina, Mari{\c
  s}, {Martinez-Soler}, Maruyama, Mase, McElroy, McNally, Mead, Meagher,
  Mechbal, Medina, Meier, {Meighen-Berger}, Micallef, Mockler, Montaruli,
  Moore, Morse, Moulai, Naab, Nagai, Nahnhauer, Naumann, Necker, Nguyen,
  Niederhausen, Nisa, Nowicki, Nygren, Obertacke~Pollmann, Oehler, Oeyen,
  Olivas, O'Sullivan, Pandya, Pankova, Park, Parker, Paudel, Paul, {P{\'e}rez
  de los Heros}, Peters, Peterson, Philippen, Pieper, Pittermann, Pizzuto,
  Plum, Popovych, Porcelli, Prado~Rodriguez, Price, Pries, Przybylski, Raab,
  {Rack-Helleis}, Raissi, Rameez, Rawlins, Rea, Rehman, Reichherzer, Reimann,
  Renzi, Resconi, Reusch, Rhode, Richman, Riedel, Roberts, Robertson,
  Roellinghoff, Rongen, Rott, Ruhe, Ryckbosch, Rysewyk~Cantu, Safa, Saffer,
  Sanchez~Herrera, Sandrock, Sandroos, Santander, Sarkar, Sarkar, Satalecka,
  Schaufel, Schieler, Schindler, Schmidt, Schneider, Schneider, Schr{\"o}der,
  Schumacher, Schwefer, Sclafani, Seckel, Seunarine, Sharma, Shefali, Silva,
  Skrzypek, Smithers, Snihur, Soedingrekso, Soldin, Spannfellner, Spiczak,
  Spiering, Stachurska, Stamatikos, Stanev, Stein, Stettner, Steuer,
  Stezelberger, Stokstad, St{\"u}rwald, Stuttard, Sullivan, Taboada,
  {Ter-Antonyan}, Tilav, Tischbein, Tollefson, T{\"o}nnis, Toscano, Tosi,
  Trettin, Tselengidou, Tung, Turcati, Turcotte, Turley, Twagirayezu, Ty,
  Unland~Elorrieta, {Valtonen-Mattila}, Vandenbroucke, {van Eijndhoven},
  Vannerom, {van Santen}, Verpoest, Walck, Watson, Weaver, Weigel, Weindl,
  Weiss, Weldert, Wendt, Werthebach, Weyrauch, Whitehorn, Wiebusch, Williams,
  Wolf, Woschnagg, Wrede, Wulff, Xu, Yanez, Yoshida, Yu, Yuan, Zhang, \&
  Zhelnin}]{icecubecollaborationEvidenceNeutrinoEmission2022}
{IceCube Collaboration}, Abbasi, R., Ackermann, M., {et~al.} 2022, Science,
  378, 538, \dodoi{10.1126/science.abg3395}

\bibitem[{Itagaki {et~al.}(2012)Itagaki, Noguchi, Nakano, Yusa, Wang, Liu,
  Zhang, \& Zhang}]{itagakiSupernova2012cwNGC2012}
Itagaki, K., Noguchi, T., Nakano, S., {et~al.} 2012, Central Bureau Electronic
  Telegrams, 3148, 1

\bibitem[{Itagaki {et~al.}(2013)Itagaki, Noguchi, Nakano, Elenin, Molotov,
  Moritani, Takaki, Kawabata, Akitaya, Ebisuda, Kawaguchi, Mori, Ohashi, Ueno,
  Sasada, Yamanaka, Ochner, Tomasella, Pastorello, Benetti, Cappellaro, \&
  Turatto}]{itagakiSupernova2013buNGC2013}
---. 2013, Central Bureau Electronic Telegrams, 3498, 1

\bibitem[{Jha {et~al.}(1999)Jha, Challis, Garnavich, Kirshner, Calkins, \&
  Stanek}]{jhaSupernova1999emNGC1999}
Jha, S., Challis, P., Garnavich, P., {et~al.} 1999, International Astronomical
  Union Circular, 7296, 2

\bibitem[{Johnson {et~al.}(2017)Johnson, Kochanek, \&
  Adams}]{johnsonProgenitorTypeIbc2017}
Johnson, S.~A., Kochanek, C.~S., \& Adams, S.~M. 2017, Monthly Notices of the
  Royal Astronomical Society, 472, 3115, \dodoi{10.1093/mnras/stx2170}

\bibitem[{Kim {et~al.}(2014)Kim, Zheng, Li, Filippenko, Cenko, Arbour, Masi,
  Nocentini, Schmeer, Zhang, Want, Tartaglia, Pastorello, Benetti, Cappellaro,
  Tomasella, Ochner, {Elias-Rosa}, \& Turatto}]{kimSupernova2014CNGC2014}
Kim, M., Zheng, W., Li, W., {et~al.} 2014, Central Bureau Electronic Telegrams,
  3777, 1

\bibitem[{Kochanek {et~al.}(2012)Kochanek, Szczygie{\l}, \&
  Stanek}]{kochanekUnmaskingSupernovaImpostors2012}
Kochanek, C.~S., Szczygie{\l}, D.~M., \& Stanek, K.~Z. 2012, The Astrophysical
  Journal, 758, 142, \dodoi{10.1088/0004-637X/758/2/142}

\bibitem[{{Kurahashi} {et~al.}(2022){Kurahashi}, {Murase}, \&
  {Santander}}]{Kurahashi:2022utm}
{Kurahashi}, N., {Murase}, K., \& {Santander}, M. 2022, arXiv e-prints,
  arXiv:2203.11936.
\newblock \doarXiv{2203.11936}

\bibitem[{Li {et~al.}(2011)Li, Leaman, {et~al.}}]{liNearbySupernovaRates2011}
Li, W., Leaman, J., {et~al.} 2011, Monthly Notices of the Royal Astronomical
  Society, 412, 1441, \dodoi{10.1111/j.1365-2966.2011.18160.x}

\bibitem[{Li {et~al.}(2005)Li, Van~Dyk, Filippenko, \&
  Cuillandre}]{liProgenitorTypeII2005}
Li, W., Van~Dyk, S.~D., Filippenko, A.~V., \& Cuillandre, J.-C. 2005,
  Publications of the Astronomical Society of the Pacific, 117, 121,
  \dodoi{10.1086/428278}

\bibitem[{Li {et~al.}(2007)Li, Wang, Van~Dyk, Cuillandre, Foley, \&
  Filippenko}]{liProgenitorsTwoType2007}
Li, W., Wang, X., Van~Dyk, S.~D., {et~al.} 2007, The Astrophysical Journal,
  661, 1013, \dodoi{10.1086/516747}

\bibitem[{Margutti {et~al.}(2013)Margutti, Soderberg, \&
  Milisavljevic}]{marguttiXrayEmissionPosition2013}
Margutti, R., Soderberg, A., \& Milisavljevic, D. 2013, The Astronomer's
  Telegram, 5106, 1

\bibitem[{Marion \& Calkins(2011)}]{marionSupernova2011anUGC2011}
Marion, G.~H., \& Calkins, M. 2011, Central Bureau Electronic Telegrams, 2668,
  2

\bibitem[{Marples \& Drescher(2009)}]{marplesSupernova2009mkPgc2009}
Marples, P., \& Drescher, C. 2009, Central Bureau Electronic Telegrams, 2080, 1

\bibitem[{{Mauerhan} {et~al.}(2013){Mauerhan}, {Smith}, {Silverman},
  {Filippenko}, {Morgan}, {Cenko}, {Ganeshalingam}, {Clubb}, {Bloom},
  {Matheson}, \& {Milne}}]{mauerhan2013}
{Mauerhan}, J.~C., {Smith}, N., {Silverman}, J.~M., {et~al.} 2013, \mnras, 431,
  2599, \dodoi{10.1093/mnras/stt360}

\bibitem[{Maxwell {et~al.}(2010)Maxwell, Graham, Parker, Sadavoy, Pritchet,
  Hsiao, \& Balam}]{maxwellSupernova2010brNGC2010}
Maxwell, A.~J., Graham, M.~L., Parker, A., {et~al.} 2010, Central Bureau
  Electronic Telegrams, 2245, 2

\bibitem[{Milisavljevic {et~al.}(2013)Milisavljevic, Fesen, Pickering,
  Miszalski, Buckley, Parrent, Marion, Silverman, Vinko, Wheeler, Quimby, Jha,
  Mohamed, Kasliwal, \&
  Soderberg}]{milisavljevicOpticalSpectroscopyIPTF13bvn2013}
Milisavljevic, D., Fesen, R., Pickering, T., {et~al.} 2013, The Astronomer's
  Telegram, 5142, 1

\bibitem[{Milisavljevic {et~al.}(2015)Milisavljevic, Margutti, Kamble,
  Patnaude, Raymond, Eldridge, Fong, Bietenholz, Challis, Chornock, Drout,
  Fransson, Fesen, Grindlay, Kirshner, Lunnan, Mackey, Miller, Parrent,
  Sanders, Soderberg, \& Zauderer}]{milisavljevicMetamorphosisSN2014C2015}
Milisavljevic, D., Margutti, R., Kamble, A., {et~al.} 2015, The Astrophysical
  Journal, 815, 120, \dodoi{10.1088/0004-637X/815/2/120}

\bibitem[{Modjaz {et~al.}(2005)Modjaz, Kirshner, Challis, \&
  Hutchins}]{modjazSupernova2005csM512005}
Modjaz, M., Kirshner, R., Challis, P., \& Hutchins, R. 2005, International
  Astronomical Union Circular, 8555, 1

\bibitem[{Monard {et~al.}(2012)Monard, Childress, Scalzo, Yuan, \&
  Schmidt}]{monardSupernova2012ecNGC2012}
Monard, L. A.~G., Childress, M., Scalzo, R., Yuan, F., \& Schmidt, B. 2012,
  Central Bureau Electronic Telegrams, 3201, 1

\bibitem[{{Moriya} {et~al.}(2011){Moriya}, {Tominaga}, {Blinnikov}, {Baklanov},
  \& {Sorokina}}]{2011MNRAS.415..199M}
{Moriya}, T., {Tominaga}, N., {Blinnikov}, S.~I., {Baklanov}, P.~V., \&
  {Sorokina}, E.~I. 2011, \mnras, 415, 199,
  \dodoi{10.1111/j.1365-2966.2011.18689.x}

\bibitem[{{Moriya} {et~al.}(2012){Moriya}, {Tominaga}, {Blinnikov}, {Baklanov},
  \& {Sorokina}}]{2012IAUS..279...54M}
{Moriya}, T.~J., {Tominaga}, N., {Blinnikov}, S.~I., {Baklanov}, P.~V., \&
  {Sorokina}, E.~I. 2012, in IAU Symposium, Vol. 279, Death of Massive Stars:
  Supernovae and Gamma-Ray Bursts, ed. P.~{Roming}, N.~{Kawai}, \& E.~{Pian},
  54--57, \dodoi{10.1017/S1743921312012689}

\bibitem[{Morrell \& Stritzinger(2008)}]{morrellSupernovae2008bk2008br2008}
Morrell, N., \& Stritzinger, M. 2008, Central Bureau Electronic Telegrams,
  1335, 1

\bibitem[{Murase {et~al.}(2011)Murase, Thompson, Lacki, \&
  Beacom}]{Murase:2010cu}
Murase, K., Thompson, T.~A., Lacki, B.~C., \& Beacom, J.~F. 2011, Phys. Rev.,
  D84, 043003, \dodoi{10.1103/PhysRevD.84.043003}

\bibitem[{Nakano {et~al.}(2013)Nakano, Kiyota, Masi, Nocentini, Schmeer, Zhang,
  \& Wang}]{nakanoSupernova2013geNGC2013}
Nakano, S., Kiyota, S., Masi, G., {et~al.} 2013, Central Bureau Electronic
  Telegrams, 3701, 1

\bibitem[{Nakaoka {et~al.}(2018)Nakaoka, Kawabata, Maeda, Tanaka, Yamanaka,
  Moriya, Tominaga, Morokuma, Takaki, Kawabata, Kawahara, Itoh, Shiki, Mori,
  Hirochi, Abe, Uemura, Yoshida, Akitaya, Moritani, Ueno, Urano, Isogai,
  Hanayama, \& Nagayama}]{nakaoka2018}
Nakaoka, T., Kawabata, K.~S., Maeda, K., {et~al.} 2018, The Astrophysical
  Journal, 859, 78.
\newblock \url{http://stacks.iop.org/0004-637X/859/i=2/a=78}

\bibitem[{Necker {et~al.}(2022)Necker, {de Jaeger}, Stein, Franckowiak,
  Shappee, Kowalski, Kochanek, Stanek, Beacom, Desai, Neumann, Jayasinghe,
  Holoien, Thompson, \& Holmbo}]{neckerASASSNFollowupIceCube2022}
Necker, J., {de Jaeger}, T., Stein, R., {et~al.} 2022, Monthly Notices of the
  Royal Astronomical Society, 516, 2455, \dodoi{10.1093/mnras/stac2261}

\bibitem[{Ochner {et~al.}(2014)Ochner, Tomasella, Benetti, Pastorello,
  {Elias-Rosa}, Cappellaro, \& Turatto}]{ochnerAsiagoClassificationPS114xz2014}
Ochner, P., Tomasella, L., Benetti, S., {et~al.} 2014, The Astronomer's
  Telegram, 6160, 1

\bibitem[{Ofek {et~al.}(2013)}]{Ofek:2013mea}
Ofek, E.~O., {et~al.} 2013, Nature, 494, 65, \dodoi{10.1038/nature11877}

\bibitem[{Parker {et~al.}(2013)Parker, Kiyota, Morrell, Hsiao, Contreras,
  Gonzalez, Campillay, Gromadzki, Ruiz, Milisavljevic, Fesen, Pickering,
  Vaisanen, Marion, Parrent, Margutti, \&
  Soderberg}]{parkerSupernova2013byESO2013}
Parker, S., Kiyota, S., Morrell, N., {et~al.} 2013, Central Bureau Electronic
  Telegrams, 3506, 1

\bibitem[{Pastorello {et~al.}(2008)Pastorello, Kasliwal, Crockett, Valenti,
  Arbour, Itagaki, Kaspi, {Gal-Yam}, Smartt, Griffith, Maguire, Ofek, Seymour,
  Stern, \& Wiethoff}]{pastorelloTypeIIbSN2008}
Pastorello, A., Kasliwal, M.~M., Crockett, R.~M., {et~al.} 2008, Monthly
  Notices of the Royal Astronomical Society, 389, 955,
  \dodoi{10.1111/j.1365-2966.2008.13618.x}

\bibitem[{Pastorello {et~al.}(2009)Pastorello, Valenti, Zampieri, Navasardyan,
  Taubenberger, Smartt, Arkharov, B{\"a}rnbantner, Barwig, Benetti,
  Birtwhistle, Botticella, Cappellaro, Del~Principe, {di Mille}, {di Rico},
  Dolci, {Elias-Rosa}, Efimova, Fiedler, Harutyunyan, H{\"o}flich, Kloehr,
  Larionov, Lorenzi, Maund, Napoleone, Ragni, Richmond, Ries, Spiro, Temporin,
  Turatto, \& Wheeler}]{pastorelloSN2005csM512009}
Pastorello, A., Valenti, S., Zampieri, L., {et~al.} 2009, Monthly Notices of
  the Royal Astronomical Society, 394, 2266,
  \dodoi{10.1111/j.1365-2966.2009.14505.x}

\bibitem[{Patat {et~al.}(2004)Patat, Benetti, Pastorello, Filippenko, \&
  Aceituno}]{patatSupernova2004djNGC2004}
Patat, F., Benetti, S., Pastorello, A., Filippenko, A.~V., \& Aceituno, J.
  2004, International Astronomical Union Circular, 8378, 1

\bibitem[{{Pitik} {et~al.}(2022){Pitik}, {Tamborra}, {Angus}, \&
  {Auchettl}}]{2022ApJ...929..163P}
{Pitik}, T., {Tamborra}, I., {Angus}, C.~R., \& {Auchettl}, K. 2022, \apj, 929,
  163, \dodoi{10.3847/1538-4357/ac5ab1}

\bibitem[{Prieto {et~al.}(2011)Prieto, McMillan, Bakos, \&
  Grennan}]{prietoSupernova2011htUGC2011}
Prieto, J.~L., McMillan, R., Bakos, G., \& Grennan, D. 2011, Central Bureau
  Electronic Telegrams, 2903, 1

\bibitem[{Quadri {et~al.}(2012)Quadri, Strabla, Girelli, Quadri, Itoh, Ui,
  Siviero, Tomasella, Pastorello, Benetti, Munari, Ergon, Sollerman, Taddia, \&
  Barisevicius}]{quadriSupernova2012awM952012}
Quadri, U., Strabla, L., Girelli, R., {et~al.} 2012, Central Bureau Electronic
  Telegrams, 3054, 1

\bibitem[{{Razzaque} {et~al.}(2004){Razzaque}, {M{\'e}sz{\'a}ros}, \&
  {Waxman}}]{2004PhRvL..93r1101R}
{Razzaque}, S., {M{\'e}sz{\'a}ros}, P., \& {Waxman}, E. 2004, Physical Review
  Letters, 93, 181101, \dodoi{10.1103/PhysRevLett.93.181101}

\bibitem[{Reusch {et~al.}(2022)Reusch, Stein, Kowalski, {van Velzen},
  Franckowiak, Lunardini, Murase, Winter, {Miller-Jones}, Kasliwal, Gilfanov,
  Garrappa, Paliya, Ahumada, Anand, Barbarino, Bellm, Brinnel, Buson, Cenko,
  Coughlin, De, Dekany, Frederick, {Gal-Yam}, Gezari, Giroletti, Graham,
  Karambelkar, Kimura, Kong, Kool, Laher, Medvedev, Necker, Nordin, Perley,
  Rigault, Rusholme, Schulze, Schweyer, Singer, Sollerman, Strotjohann,
  Sunyaev, {van Santen}, Walters, Zhang, \&
  Zimmerman}]{reuschCandidateTidalDisruption2022a}
Reusch, S., Stein, R., Kowalski, M., {et~al.} 2022, Physical Review Letters,
  128, 221101, \dodoi{10.1103/PhysRevLett.128.221101}

\bibitem[{Sarmah {et~al.}(2022)Sarmah, Chakraborty, Tamborra, \&
  Auchettl}]{sarmahHighEnergyParticles2022}
Sarmah, P., Chakraborty, S., Tamborra, I., \& Auchettl, K. 2022, Journal of
  Cosmology and Astroparticle Physics, 2022, 011,
  \dodoi{10.1088/1475-7516/2022/08/011}

\bibitem[{Senno {et~al.}(2016)Senno, Murase, \& Meszaros}]{Senno:2015tsn}
Senno, N., Murase, K., \& Meszaros, P. 2016, Phys. Rev., D93, 083003,
  \dodoi{10.1103/PhysRevD.93.083003}

\bibitem[{Shivvers {et~al.}(2014)Shivvers, Kelly, Clubb, \&
  Filippenko}]{shivversSpectroscopicClassificationMASTER2014}
Shivvers, I., Kelly, P.~L., Clubb, K.~I., \& Filippenko, A.~V. 2014, The
  Astronomer's Telegram, 6487, 1

\bibitem[{Silverman {et~al.}(2008)Silverman, Griffith, Filippenko, Chornock, \&
  Li}]{silvermanSupernovae2008dv2008dz2008}
Silverman, J.~M., Griffith, C.~V., Filippenko, A.~V., Chornock, R., \& Li, W.
  2008, Central Bureau Electronic Telegrams, 1447, 1

\bibitem[{Singh {et~al.}(2019)Singh, Misra, Sahu, Dastidar, Gangopadhyay,
  Srivastav, Anupama, Bose, Lipunov, Chakradhari, Kumar, Kumar, Pandey,
  Gorbovskoy, \& Balanutsa}]{singhObservationalPropertiesType2019}
Singh, M., Misra, K., Sahu, D.~K., {et~al.} 2019, Monthly Notices of the Royal
  Astronomical Society, 485, 5438, \dodoi{10.1093/mnras/stz752}

\bibitem[{Smith {et~al.}(2011)Smith, Li, Silverman, Ganeshalingam, \&
  Filippenko}]{smithLuminousBlueVariable2011a}
Smith, N., Li, W., Silverman, J.~M., Ganeshalingam, M., \& Filippenko, A.~V.
  2011, Monthly Notices of the Royal Astronomical Society, 415, 773,
  \dodoi{10.1111/j.1365-2966.2011.18763.x}

\bibitem[{Sollerman {et~al.}(2009)Sollerman, Ergon, Inserra, Valenti, Wilson,
  Jon~Juliusson, Holma, Ingemyr, Saxen, \&
  Haukanes}]{sollermanSupernova2009mdNGC2009}
Sollerman, J., Ergon, M., Inserra, C., {et~al.} 2009, Central Bureau Electronic
  Telegrams, 2068, 1

\bibitem[{Srivastav {et~al.}(2014{\natexlab{a}})Srivastav, Anupama, \&
  Sahu}]{srivastavOpticalObservationsFast2014}
Srivastav, S., Anupama, G.~C., \& Sahu, D.~K. 2014{\natexlab{a}}, Monthly
  Notices of the Royal Astronomical Society, 445, 1932,
  \dodoi{10.1093/mnras/stu1878}

\bibitem[{Srivastav {et~al.}(2014{\natexlab{b}})Srivastav, Sahu, \&
  Anupama}]{srivastavSpectroscopicClassificationMASTER2014}
Srivastav, S., Sahu, D.~K., \& Anupama, G.~C. 2014{\natexlab{b}}, The
  Astronomer's Telegram, 6639, 1

\bibitem[{Stanishev {et~al.}(2008)Stanishev, Pastorello, \&
  Pursimo}]{stanishevSupernova2008SNGC2008}
Stanishev, V., Pastorello, A., \& Pursimo, T. 2008, Central Bureau Electronic
  Telegrams, 1235, 1

\bibitem[{Stanishev \& Pursimo(2012)}]{stanishevSupernova2012ANGC2012}
Stanishev, V., \& Pursimo, T. 2012, Central Bureau Electronic Telegrams, 2974,
  3

\bibitem[{Steele {et~al.}(2009{\natexlab{a}})Steele, Cobb, \&
  Filippenko}]{steeleSupernovae2009kn2009ko2009}
Steele, T.~N., Cobb, B., \& Filippenko, A.~V. 2009{\natexlab{a}}, Central
  Bureau Electronic Telegrams, 2011, 1

\bibitem[{Steele {et~al.}(2009{\natexlab{b}})Steele, Kandrashoff, \&
  Filippenko}]{steeleSupernovae2009lu2009md2009}
Steele, T.~N., Kandrashoff, M.~T., \& Filippenko, A.~V. 2009{\natexlab{b}},
  Central Bureau Electronic Telegrams, 2070, 1

\bibitem[{Stein(2019)}]{Stein:2019lI}
Stein, R. 2019, in Proceedings of 36th International Cosmic Ray Conference
  {\textemdash} PoS(ICRC2019), Vol. 358, 1016, \dodoi{10.22323/1.358.1016}

\bibitem[{Stein {et~al.}(2022{\natexlab{a}})Stein, Necker, Bradascio, \&
  Garrappa}]{flarestack}
Stein, R., Necker, J., Bradascio, F., \& Garrappa, S. 2022{\natexlab{a}},
  icecube/flarestack: Titan v2.4.2, v2.4.2,  Zenodo,
  \dodoi{10.5281/zenodo.6425740}

\bibitem[{Stein {et~al.}(2021)}]{Stein:2020xhk}
Stein, R., {et~al.} 2021, Nature Astron., 5, 510,
  \dodoi{10.1038/s41550-020-01295-8}

\bibitem[{Stein {et~al.}(2022{\natexlab{b}})Stein, Reusch, Franckowiak,
  Kowalski, Necker, Weimann, Kasliwal, Sollerman, Ahumada, Seoane, Anand,
  Andreoni, Bellm, Bloom, Coughlin, De, Fremling, Gezari, Graham, Groom, Helou,
  Kaplan, Karambelkar, Kong, Kool, Lincetto, Mahabal, Masci, Medford, Morgan,
  Nordin, Rodriguez, Sharma, {van Santen}, {van Velzen}, \&
  Yan}]{steinNeutrinoFollowupZwicky2022}
Stein, R., Reusch, S., Franckowiak, A., {et~al.} 2022{\natexlab{b}}, arXiv
  e-prints, arXiv:2203.17135

\bibitem[{Stockdale {et~al.}(2009)Stockdale, Rentz, Vandrevala, Weiler, Immler,
  {van Dyk}, Panagia, Marcaide, Pooley, Sramek, \&
  Ryder}]{stockdaleSupernova2009gjNGC2009}
Stockdale, C.~J., Rentz, B., Vandrevala, C.~M., {et~al.} 2009, International
  Astronomical Union Circular, 9056, 1

\bibitem[{Stritzinger {et~al.}(2010)Stritzinger, Folatelli, \&
  Pignata}]{stritzingerSupernova2009muESO2010}
Stritzinger, M., Folatelli, G., \& Pignata, G. 2010, Central Bureau Electronic
  Telegrams, 2116, 1

\bibitem[{Strolger {et~al.}(2015)Strolger, Dahlen, Rodney, Graur, Riess,
  McCully, Ravindranath, Mobasher, \& Shahady}]{Strolger:2015kra}
Strolger, L.-G., Dahlen, T., Rodney, S.~A., {et~al.} 2015, Astrophys. J., 813,
  93, \dodoi{10.1088/0004-637X/813/2/93}

\bibitem[{{Strotjohann} {et~al.}(2021){Strotjohann}, {Ofek}, {Gal-Yam},
  {Bruch}, {Schulze}, {Shaviv}, {Sollerman}, {Filippenko}, {Yaron}, {Fremling},
  {Nordin}, {Kool}, {Perley}, {Ho}, {Yang}, {Yao}, {Soumagnac}, {Graham},
  {Barbarino}, {Tartaglia}, {De}, {Goldstein}, {Cook}, {Brink}, {Taggart},
  {Yan}, {Lunnan}, {Kasliwal}, {Kulkarni}, {Nugent}, {Masci}, {Rosnet},
  {Adams}, {Andreoni}, {Bagdasaryan}, {Bellm}, {Burdge}, {Duev}, {Dugas},
  {Frederick}, {Goldwasser}, {Hankins}, {Irani}, {Karambelkar}, {Kupfer},
  {Liang}, {Neill}, {Porter}, {Riddle}, {Sharma}, {Short}, {Taddia},
  {Tzanidakis}, {van Roestel}, {Walters}, \& {Zhuang}}]{2021ApJ...907...99S}
{Strotjohann}, N.~L., {Ofek}, E.~O., {Gal-Yam}, A., {et~al.} 2021, \apj, 907,
  99, \dodoi{10.3847/1538-4357/abd032}

\bibitem[{Szczygie{\l} {et~al.}(2012)Szczygie{\l}, Kochanek, \&
  Dai}]{szczygielSN2002buAnother2012}
Szczygie{\l}, D.~M., Kochanek, C.~S., \& Dai, X. 2012, The Astrophysical
  Journal, 760, 20, \dodoi{10.1088/0004-637X/760/1/20}

\bibitem[{Takaki {et~al.}(2012)Takaki, Itoh, Ueno, Urano, Moritani, Akitaya,
  Kawabata, \& Yamanaka}]{takakiSupernova2012fhNGC2012}
Takaki, K., Itoh, R., Ueno, I., {et~al.} 2012, Central Bureau Electronic
  Telegrams, 3263, 3

\bibitem[{Tartaglia {et~al.}(2020)Tartaglia, Pastorello, Sollerman, Fransson,
  Mattila, Fraser, Taddia, Tomasella, Turatto, {Morales-Garoffolo},
  {Elias-Rosa}, Lundqvist, Harmanen, Reynolds, Cappellaro, Barbarino, Nyholm,
  Kool, Ofek, Gao, Jin, Tan, Sand, Ciabattari, Wang, Zhang, Huang, Li, Mo, Rui,
  Xiang, Zhang, Hosseinzadeh, Howell, McCully, Valenti, Benetti, Callis,
  Carracedo, Fremling, Kangas, Rubin, Somero, \&
  Terreran}]{tartagliaLonglivedTypeIIn2020}
Tartaglia, L., Pastorello, A., Sollerman, J., {et~al.} 2020, Astronomy and
  Astrophysics, 635, A39, \dodoi{10.1051/0004-6361/201936553}

\bibitem[{Tinyanont {et~al.}(2016)Tinyanont, Kasliwal, Fox, Lau, Smith,
  Williams, Jencson, Perley, Dykhoff, Gehrz, Johansson, Van~Dyk, Masci, Cody,
  \& Prince}]{tinyanontSystematicStudyMidinfrared2016}
Tinyanont, S., Kasliwal, M.~M., Fox, O.~D., {et~al.} 2016, The Astrophysical
  Journal, 833, 231, \dodoi{10.3847/1538-4357/833/2/231}

\bibitem[{Tomasella {et~al.}(2012)Tomasella, Turatto, Benetti, Pastorello,
  Ochner, \& Cappellaro}]{tomasellaSupernova2012fhNGC2012}
Tomasella, L., Turatto, M., Benetti, S., {et~al.} 2012, Central Bureau
  Electronic Telegrams, 3263, 2

\bibitem[{Valenti \& Benetti(2011)}]{valentiSupernova2011dqNGC2011}
Valenti, S., \& Benetti, S. 2011, Central Bureau Electronic Telegrams, 2749, 2

\bibitem[{Valenti {et~al.}(2008)Valenti, {Elias-Rosa}, Taubenberger, Stanishev,
  Agnoletto, Sauer, Cappellaro, Pastorello, Benetti, Riffeser, Hopp,
  Navasardyan, Tsvetkov, Lorenzi, Patat, Turatto, Barbon, Ciroi, Di~Mille,
  Frandsen, Fynbo, Laursen, \& Mazzali}]{valentiCarbonrichTypeIc2008}
Valenti, S., {Elias-Rosa}, N., Taubenberger, S., {et~al.} 2008, The
  Astrophysical Journal, 673, L155, \dodoi{10.1086/527672}

\bibitem[{Van~Dyk {et~al.}(2014)Van~Dyk, Zheng, Fox, Cenko, Clubb, Filippenko,
  Foley, Miller, Smith, Kelly, Lee, {Ben-Ami}, \&
  {Gal-Yam}}]{vandykTypeIIbSupernova2014}
Van~Dyk, S.~D., Zheng, W., Fox, O.~D., {et~al.} 2014, The Astronomical Journal,
  147, 37, \dodoi{10.1088/0004-6256/147/2/37}

\bibitem[{Wang {et~al.}(2012)Wang, Liu, Zhang, Zhang, \&
  Brimacombe}]{wangSupernova2012cwNGC2012}
Wang, X.~F., Liu, Q., Zhang, J.~J., Zhang, T.~M., \& Brimacombe, J. 2012,
  Central Bureau Electronic Telegrams, 3148, 2

\bibitem[{Yamanaka {et~al.}(2010)Yamanaka, Arai, Sakimoto, Okushima, \&
  Kawabata}]{yamanakaSupernova2010giIC2010}
Yamanaka, M., Arai, A., Sakimoto, K., Okushima, T., \& Kawabata, K.~S. 2010,
  Central Bureau Electronic Telegrams, 2384, 1

\bibitem[{Yaron \& Gal-Yam(2012)}]{Yaron:2012aa}
Yaron, O., \& Gal-Yam, A. 2012, Publications of the Astronomical Society of the
  Pacific, 124, 668.
\newblock \url{http://stacks.iop.org/1538-3873/124/i=917/a=668}

\bibitem[{Yaron {et~al.}(2017)}]{Yaron:2017umb}
Yaron, O., {et~al.} 2017, Nature Phys., 13, 510, \dodoi{10.1038/nphys4025}

\bibitem[{Zhang \& Wang(2015)}]{zhangSpectroscopicClassificationPSN2015}
Zhang, J., \& Wang, X. 2015, The Astronomer's Telegram, 6939, 1

\bibitem[{Zirakashvili \& Ptuskin(2016)}]{Zirakashvili:2015mua}
Zirakashvili, V.~N., \& Ptuskin, V.~S. 2016, Astropart. Phys., 78, 28,
  \dodoi{10.1016/j.astropartphys.2016.02.004}

\bibitem[{Zwitter {et~al.}(2004)Zwitter, Munari, \&
  Moretti}]{zwitterSupernova2004etNGC2004}
Zwitter, T., Munari, U., \& Moretti, S. 2004, International Astronomical Union
  Circular, 8413, 1

\end{thebibliography}
